\documentclass{aer} 

\usepackage[figuresleft]{rotating}

\usepackage{float}
\usepackage[hidelinks]{hyperref}
\usepackage{hyperref}
\usepackage{amsmath}
\usepackage{subcaption}
\usepackage{multirow}
\usepackage[justification=centering]{caption}
\usepackage{booktabs}
\usepackage{tabularx}
\usepackage{adjustbox}
\usepackage{lineno}
\usepackage{txfonts}%
\usepackage{titlesec}
\titleclass{\subsubsubsection}{straight}[\subsubsection]
\newcounter{subsubsubsection}[subsubsection]
\renewcommand\thesubsubsubsection{\thesubsubsection.\arabic{subsubsubsection}}

\titleformat{\section}{\normalfont\Large\bfseries}{\thesection}{1em}{}
\titlespacing*{\section}{0pt}{2ex plus 1ex minus .2ex}{1ex plus .2ex}

\titleformat{\subsection}{\normalfont\large\bfseries}{\thesubsection}{1em}{}
\titlespacing*{\subsection}{0pt}{2ex plus 1ex minus .2ex}{1ex plus .2ex}

\titleformat{\subsubsection}{\normalfont\large\bfseries}{\thesubsubsection}{1em}{}
\titlespacing*{\subsubsection}{0pt}{3.25ex plus 1ex minus .2ex}{1.5ex plus .2ex}

\titleformat{\subsubsubsection}{\normalfont\normalsize\bfseries}{\thesubsubsubsection}{1em}{}
\titlespacing*{\subsubsubsection}{0pt}{3.25ex plus 1ex minus .2ex}{1.5ex plus .2ex}

\jmonth{Month}
\jyear{YYYY}
\volume{XXX}
\issue{xxx}
\historydate{Received DD MM YYYY; revised DD MM YYYY; accepted DD MM YYYY}
\doi{xx.xxxx/aer.xxxx.x} 

\begin{document}
\markboth{Fraihat and Ajaj}{Aeroelastic Tailoring \ldots}


\title{Aeroelastic tailoring of stiffened cantilever plate using composites and structural layouts: A parametric study} 


\author{Bara’ah Fraihat and Rafic M. Ajaj \thanks{{\bf Author's notes}} \email{Corresponding author: R. M.  Ajaj;  \href{mailto:rafic.ajaj@ku.ac.ae}{rafic.ajaj@ku.ac.ae}; Tel.: +971-(0)-2-3124238}}
\affil{Khalifa University\\ Department of Aerospace Engineering\\ Abu Dhabi 127788\\ United Arab Emirates}

\maketitle                   
\begin{abstract}
This paper studies the aeroelasticity of a stiffened cantilever plate using composite material and novel structural layouts. A comprehensive parametric study is conducted to determine the influence of different design parameters on the aeroelastic boundaries. Design parameters include plate sweep angle, ply orientation, stringer cross-section, and stringer sweep angle. Nastran is used to run the aeroelastic analysis and the process is automated using $\mathrm{Matlab^{TM}}$. The structure of the plate is modelled using laminate elements whereas the stringers are modelled using the Euler-Bernoulli beam elements. The unsteady aerodynamic loads are modelled using Doublet Lattice Method (DLM) and the structural and aerodynamic meshes are connected using an Infinite plate surface (IPS) spline. A mesh sensitivity analysis is conducted to ensure fine meshes for the aerodynamics and structure. The study's findings demonstrate the benefits of employing forward swept (Fw) stringers since it increases flutter speed by almost 38\% compared to the unswept stringers case and prevents divergence. Moreover, the static aeroelastic analysis illustrates that the utilization of Fw swept stringers can reduce the average tip displacement and tip twist effectively.  T-shaped stringers are recommended to stiffen the plate due to their lower impact on the total mass of the plate. In some configurations, the structural layout has a much higher effect on the aeroelastic instabilities when compared to the material effect (ply orientation). However, results suggest combining both for some cases to get balanced washin and washout effects.  
\end{abstract}




\section{INTRODUCTION}
To fulfil the ambitious objectives outlined in the FlightPath2050 document \cite{ref1} in reducing fuel burn, noise, and emissions, aircraft manufacturers have been designing and building aircraft with higher aspect ratio wings. High aspect ratio wings play a pivotal role in reducing drag (mainly induced drag), allowing improved flight performance and extended range/endurance. These high aspect ratio wings are usually very flexible, making them prone to aeroelastic instabilities and structural failures when encountering gusts during flight; hence, methods to control these instabilities have always been a critical part of aeronautical engineering \cite{ref2}. In 2003, turbulence caused the Helios (HP03) flying wing, to bend into a high dihedral configuration. This occurred as a consequence of the wing's remarkable flexibility, leading to an unstable pitch mode. The speed of the Helios exceeded the designed airspeed which in turn increased the dynamic pressure and led to a failure of the wing leading edge secondary structure on the outer wing panels. Additionally, it resulted in the detachment of the solar cell skins on the upper surface of the wing. One of the problems that the Helios incident represented is the complex interactions between the flexible structure and unsteady aerodynamics \cite{ref3} known as aeroelasticity. 
Aeroelasticity is the discipline that addresses the interaction of aerodynamic, elastic, and inertia (including gravitational) forces. Like the Euler strut under end load, aeroelasticity is concerned with stiffness, not strength \cite{ref4}. According to Shirk et al. \cite{ref5}, aeroelastic tailoring can be defined as “the embodiment of directional stiffness into an aircraft structural design to control aeroelastic deformation, static or dynamic, in such a fashion as to affect the aerodynamic and structural performance of that aircraft in a beneficial way” and it is one of the methods to favourably use the bending-torsion coupling of the flexible wing to control aeroelastic instabilities such as flutter and divergence. Another definition was provided by Weisshaar et al. \cite{ref6} who stated that in theory, wing structural tailoring may be defined as an adjustment of the primary stiffness axis of the wing to improve aeroelastic performance. In 1949, Munk \cite{ref5} was the first to apply the concept of aeroelastic tailoring by orienting the grain (fibres) of his wooden propeller blade to create desirable deformation couplings as the load increases during operation. The development of composite materials in the 1970s ushered in a new era of aircraft design, enabling the design of rugged airframes and structures which are stiffer than those constructed of conventional materials while remaining lightweight and able to withstand aerodynamic forces. The X-29, a Forward swept (Fw) wing aircraft, was the first to apply the concept of using advanced composites in aeroelastic tailoring. The need for minimum weight to improve aircraft performance is the primary design driver for most modern aircraft. The use of composite materials in the aircraft structure has proven to be beneficial in achieving optimum performance with minimum weight. This gain is mainly due to the directional stiffness, strength properties, and high stiffness-to-weight ratios of composites \cite{ref7}.  Historically, the use of composites in wing structures is arguably the most common type of aeroelastic tailoring. Some of the composite tailoring studies focused on modelling the wing as an anisotropic plate \cite{ref8, ref9, ref10} or beam \cite{ref11, ref12} since it is faster to analyse, meanwhile, it gives an accurate estimation of how the aeroelastic instabilities change with varying ply orientation.  Sherrer et al. \cite{ref9} concluded that composite plate wings, compared to aluminium plate wings, were more effective in preventing divergence. Sherrer also suggested that divergence speed can be altered by changing the orientation of the composite laminate about the wing’s structural reference line. Green \cite{ref13} studied the aeroelastic performance of backward swept (Aft) high-aspect-ratio wings. In the study, symmetric laminates showed several advantages over general/nonsymmetric laminates, e.g. (no warping of the laminate, easier to analyse and offer fewer design decisions). General laminates introduced two additional coupling parameters to the analysis; (1) extension-torsion coupling, which led to a flutter boundary degradation, and (2) extension-bending coupling, which was less damaging than the extension-torsion coupling and gave a decent flutter boundary compared to the best symmetric laminate found in the work. Furthermore, when both parameters were introduced in the analysis, the performance of the wing was degraded compared to the reference symmetric laminate. Generally, composite tailoring studies evaluated the effect of varying the ply orientation on the structure’s flexural rigidity (EI) and torsional stiffness (GJ) and concluded that an increase in the flutter speed is associated with an increase in GJ. Moreover, the work done by \cite{ref10} and \cite{ref12} concluded that aeroelastic response can show some discontinuities (e.g., a rapid change in flutter speed associated with mode change and frequency peaks) when slightly altering the laminate ply orientation. Other methods of aeroelastic tailoring focus on the wing's structural arrangements and geometrical layouts. Harmin et al. \cite{ref14} proved the possibility of metallic aeroelastic tailoring by varying the rib’s orientation and by making use of crenelations in the wing skin. Both affected the bending torsion coupling and showed a 3\% increase in the flutter speed. Locatelli et al. \cite{ref15} used curvilinear spars and ribs (SpaRibs) to reduce torsional deformation due to aerodynamic loads thus bending-torsion coupling can be exploited to maximise flutter speed. Francois et al. \cite{ref16} investigated the influence of rib/spar arrangement on aeroelastic performance using both finite element modelling (FEM) and experimental testing. Results didn’t show a full agreement between both methods; however, both methods found that modifying the orientation of the ribs altered the structural bend-twist coupling, consequently influencing the wing's behaviour under static and aeroelastic loading. These alterations were found to be correlated with a shift in the location of the flexural axis. Passive Aeroelastic Tailoring (PAT) as referred to in \cite{ref17} is an example of a recent project to explore innovative approaches to accomplish aeroelastic tailoring on high aspect ratio wings. Passive methods of tailoring do not require integrating external sensors and actuators into the structure and aerodynamics, hence, maintaining minimum payloads and avoiding issues associated with control surfaces. Tailoring the material and arrangement of the main load-bearing elements in the wing box, specifically the skin and stringers, can lead to a substantial coupling effect between bending and twisting loads.  
This paper presents a comprehensive parametric study on the aeroelasticity of a stiffened composite cantilever flat plat. Design parameters include plate sweep angle, ply orientation, stringer cross-section, and stringer sweep angle. The wing model is approximated as a multi-layered composite plate, clamped at the root, and stiffened by integral stringers (modelled using beam element). The aerodynamic loads are estimated using Doublet Lattice Method (DLM). The aeroelastic analysis is performed using Nastran. The simulation process is automated using $\mathrm{Matlab^{TM}}$.

\section{Aeroelastic modelling}

\subsection{Geometrical configuration}
\begin{figure}[H]
\begin{center}
\includegraphics[width=3.5in,height=2.5in]{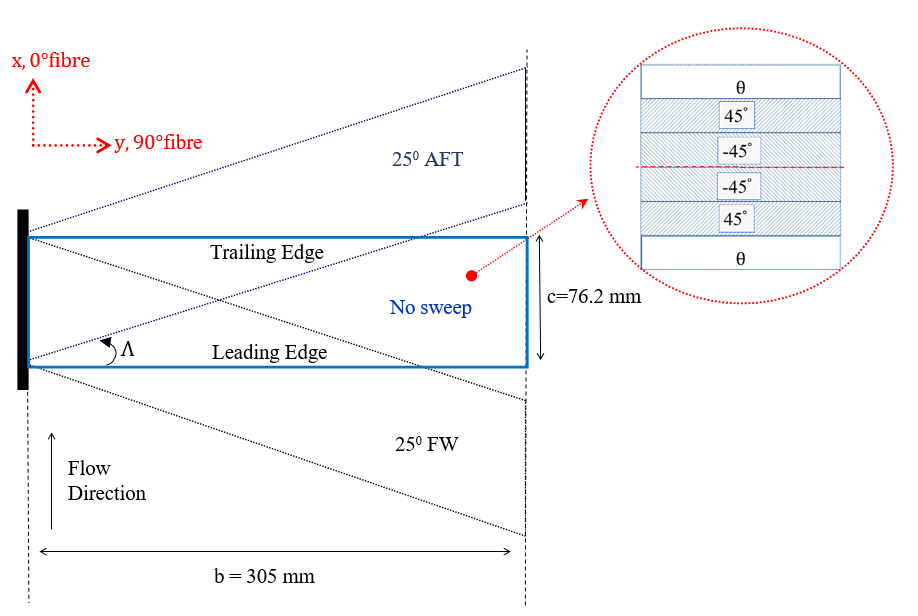}
\end{center}
\caption{Baseline plate geometry and fibre convention.}
\label{fig:fig1}
\end{figure}

This study examines the aeroelastic behaviour of a composite flat plate through two primary configurations. The first configuration serves as the baseline, consisting of a plate without stringers as shown in Figure \ref{fig:fig1}. Three different sweep angles, $\Lambda$, are considered (unswept, 25 degrees Fw sweep, and 25 degrees Aft sweep). Regardless of the sweep angle, the aerodynamic span is 305 mm. The mass of the multi-layered unswept plate is approximately 0.0284 kg. The fibre angle convention used is defined as 0° fibre when parallel to the flow velocity vector (x-axis) from the leading edge to the trailing edge, and 90° fibre travels along the semispan (y-axis) from root to tip. Throughout the entire work, only the first and last layers, theta, $\theta$, of the laminate are varied in the sequence of $[\theta,45^\circ,-45^\circ]_s$. This selection is made since the first and last layers are the furthest from the neutral axis, thus they contribute the most to bending rigidity. 
The Graphite/Epoxy plate material and geometrical properties are taken from \cite{ref18, ref19}. Table \ref{tab1} summarises the laminate properties.
\begin{table}[h]
\tabcolsep7.5pt
\caption{laminate properties \cite{ref19}}
\label{tab1}
\centering
\begin{tabular}{@{}c|c|c@{}}
\hline\hline

Ply stacking sequence (deg) & Number of plies & Ply thickness (mm) \\\hline
$[\theta,45^\circ,-45^\circ]_s$ & 6 &  0.134  \\\hline\hline

\end{tabular}
\end{table}

\begin{figure}[H]
\begin{center}
\includegraphics[width=2.6in,height=0.9in]{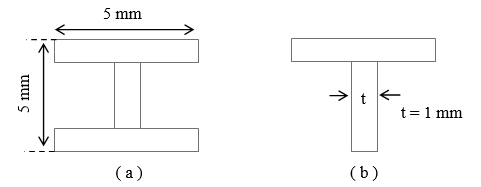}
\end{center}
\caption{(a) I (b) T-shaped stringers.}
\label{fig:fig2}
\end{figure}

In the second configuration, an unswept, stiffened plate is considered. This involves using I and T-shaped stringers to enhance the stiffness of the plate. Initially, the stringers are placed parallel to the structural reference axis of the plate. Then, the orientation of the stringers is modified by introducing an angle between the stringers and the structural reference axis of the plate, referred to as swept stringers.  In both cases, the spacing between the stringers is kept constant. This means that the distance between stringers remains unchanged regardless of whether the stringers are swept or unswept relative to the plate. Additionally, the mass of the plate is consistent for each cross-section of the stringers, regardless of their orientation (swept or unswept). The stringers are made from Aluminium (AL 2024-T3) whose properties are taken from \cite{ref20}. Two different cross-sections are used as shown in Figure \ref{fig:fig2}, both with the same height, length, and thickness. 

\begin{figure}[h]
\begin{center}
\includegraphics[width=3.4in,height=1.75in]{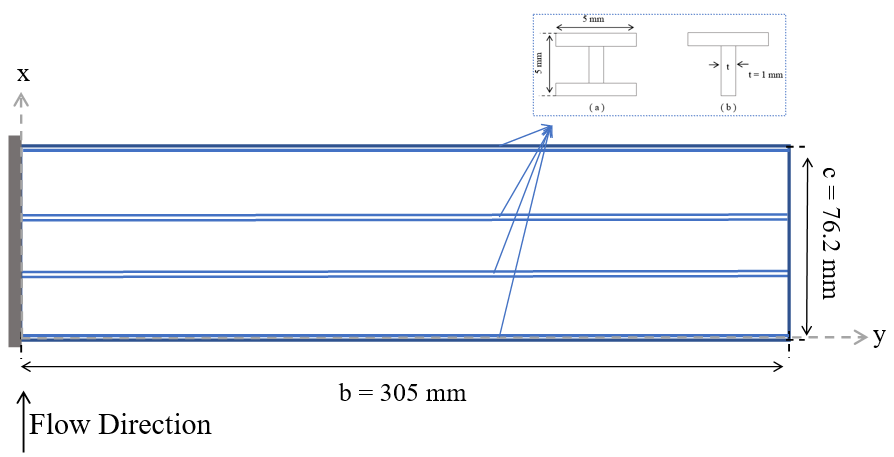}
\end{center}
\caption{Unswept stiffened plate.}
\label{fig:fig3}
\end{figure}

\begin{figure}[h]
\begin{center}
\includegraphics[width=2.5in,height=0.5in]{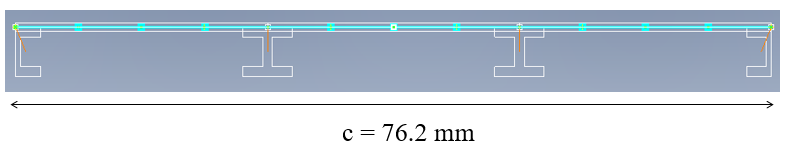}
\end{center}
\caption{A side view of the composite plate stiffened with I-shaped stringers.}
\label{fig:fig4}
\end{figure}

Figure \ref{fig:fig3} illustrates the unswept composite stiffened plate, where the stringers are placed parallel to the plate’s structural reference axis.
To ensure consistency in the total mass of the stiffened plate and maintain the stringer spacing for both the unswept and swept stringer cases, a modification is made specifically for the unswept stringers case. In this modification, the cross-section of the trailing edge and leading-edge stringers is halved compared to the actual stringer cross-section as shown in Figure \ref{fig:fig4} for a plate stiffened with I-shaped stringers.

\begin{figure}[H]
\begin{center}
\includegraphics[width=3.4in,height=1.7in]{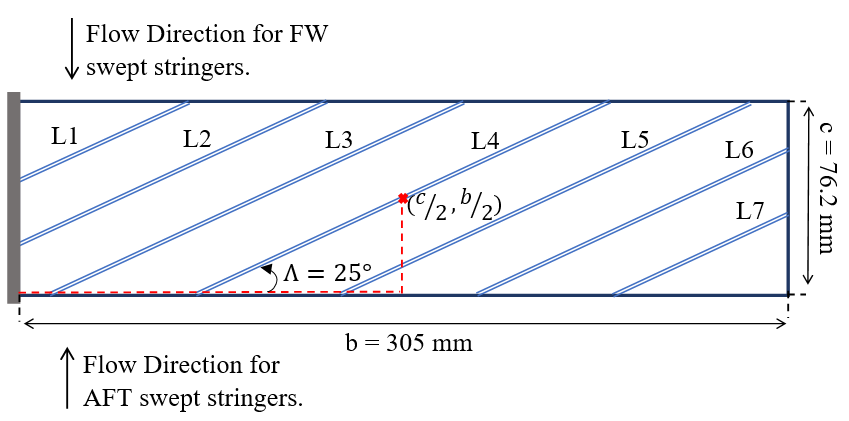}
\end{center}
\caption{Unswept stiffened plate with swept stringers. The figure illustrates two flow directions; one to analyse the effect of sweeping the stringers Aft (positive $\theta$), and the other when sweeping them Fw (negative $\theta$).}
\label{fig:fig5}
\end{figure}

\noindent Figure \ref{fig:fig5} shows the unswept stiffened plate with swept stringers. Table \ref{tab2} illustrate the length of each swept stringer on the plate. The swept stringers are distributed with equal spacing around the centre of gravity (CG) so that the locus of CG is kept fixed. However, each strip (airfoil) of the plate has its own CG. 

\begin{table}[H]
\tabcolsep7.5pt
\caption{Length of each swept stringer illustrated in Figure \ref{fig:fig4}}

\label{tab2}
\centering
\begin{tabular}{@{}c|c|c|c@{}}
\hline\hline

\multicolumn{4}{c}{{Length of each stringer (m)}}\\\hline\hline
L1 = L7 & L2 = L6 & L3 = L5 & L4  \\\hline
0.0751269 & 0.135228 & 0.180305 & 0.180304 \\\hline\hline

\end{tabular}
\end{table}

\subsection{Flexural Stiffness of the Graphite/Epoxy Plate}

\begin{figure}[h]
\begin{center}
\includegraphics[width=3.25in,height=2.25in]{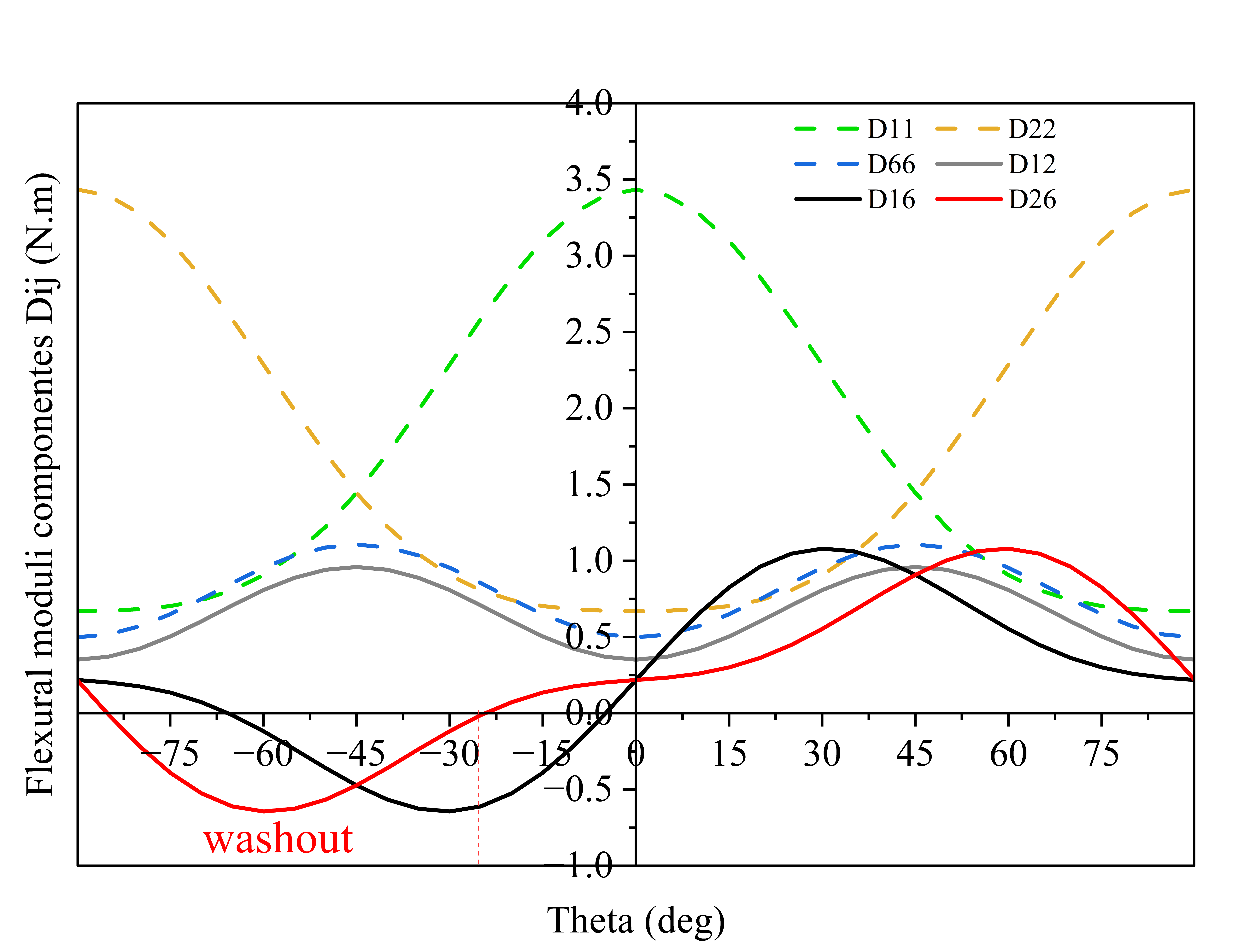}
\end{center}
\caption{Flexural moduli components for $-90^\circ<\mathrm{\theta}<90^\circ$.}
\label{fig:fig6}
\end{figure}

$\mathbf{ABD} \in \mathbb{R}^{6\times6}$ relates the applied load with the associated strains in the laminate. \textbf{A}, \textbf{B}, and \textbf{D} are the extensional stiffness, bending-extension coupling, and flexural stiffness matrices respectively. 
\begin{equation}
    \begin{Bmatrix}
     \mathbf{N}\\ \mathbf{M}\end{Bmatrix} = \begin{bmatrix}
        \mathbf{A} & \mathbf{B}\\
        \mathbf{B} & \mathbf{D}
    \end{bmatrix}\begin{Bmatrix}
        \epsilon \\\kappa
    \end{Bmatrix}
\end{equation}where \textbf{N} is the vector of in-plane loads, \textbf{M} is the vector of bending/twisting moments, and $\mathrm{\epsilon}$ and $\kappa$ are the resulting mid-plane strains and curvatures respectively. From an aeroelastic perspective, composite-based tailoring can positively or negatively alter the coupling terms in the \textbf{ABD}. The laminate stacking sequence is one of the main design drivers. The coupling terms $\mathrm{B}_{16}$ (bending-extension) and $\mathrm{B}_{26}$ (torsion-extension) which can be found in \textbf{B}, are an example of undesirable coupling, where the transverse loads along a wing cause both typical bending curvatures and an atypical in-plane extension, which can be large and nonlinear \cite{ref19}. Nevertheless, in this paper, a symmetric staking sequence was used to eliminate the possibility of having any unfavourable coupling caused by \textbf{B}. Altering \textbf{D} coupling terms ($\mathrm{D}_{16}$) and ($\mathrm{D}_{26}$) is the most well-known and used strategy for tailoring by composites. The flexural modulus $\mathrm{D}_{ij}$ for n-ply laminate with arbitrary ply angle orientation can be obtained using: 
\begin{equation}
    \mathrm{D}_{ij} = \frac{1}{3}\sum_{k=1}^n \mathrm{Q}_{ij}^{(\theta_k)}\left[z_k^3 - z_{k-1}^3\right] \qquad \qquad ij = 1,2...6
\end{equation}
where $\mathrm{Q}_{ij}^{(\theta_k)}$ is the off-axis lamina modulus of the $\mathrm{k}^{th}$ ply, $\mathrm{\theta}_k$ is the ply angle of the $\mathrm{k}^{th}$ ply and $\mathrm{z_k}$ is the vertical distance from the midplane to the upper surface of the $\mathrm{k^{th}}$ ply \cite{ref18}. The flexural moduli components $\mathrm{D}_{ij}$ for the stacking sequence used in this work ($[\mathrm{\theta},45^\circ,-45^\circ]_s$) on the entire range of theta ($-90^\circ<\mathrm{\theta}<90^\circ$) were computed using Nastran and plotted in  Figure \ref{fig:fig6}.

\subsection{Structural Model}
\begin{figure}[H]
\begin{center}
\includegraphics[width=3.55in,height=2.35in]{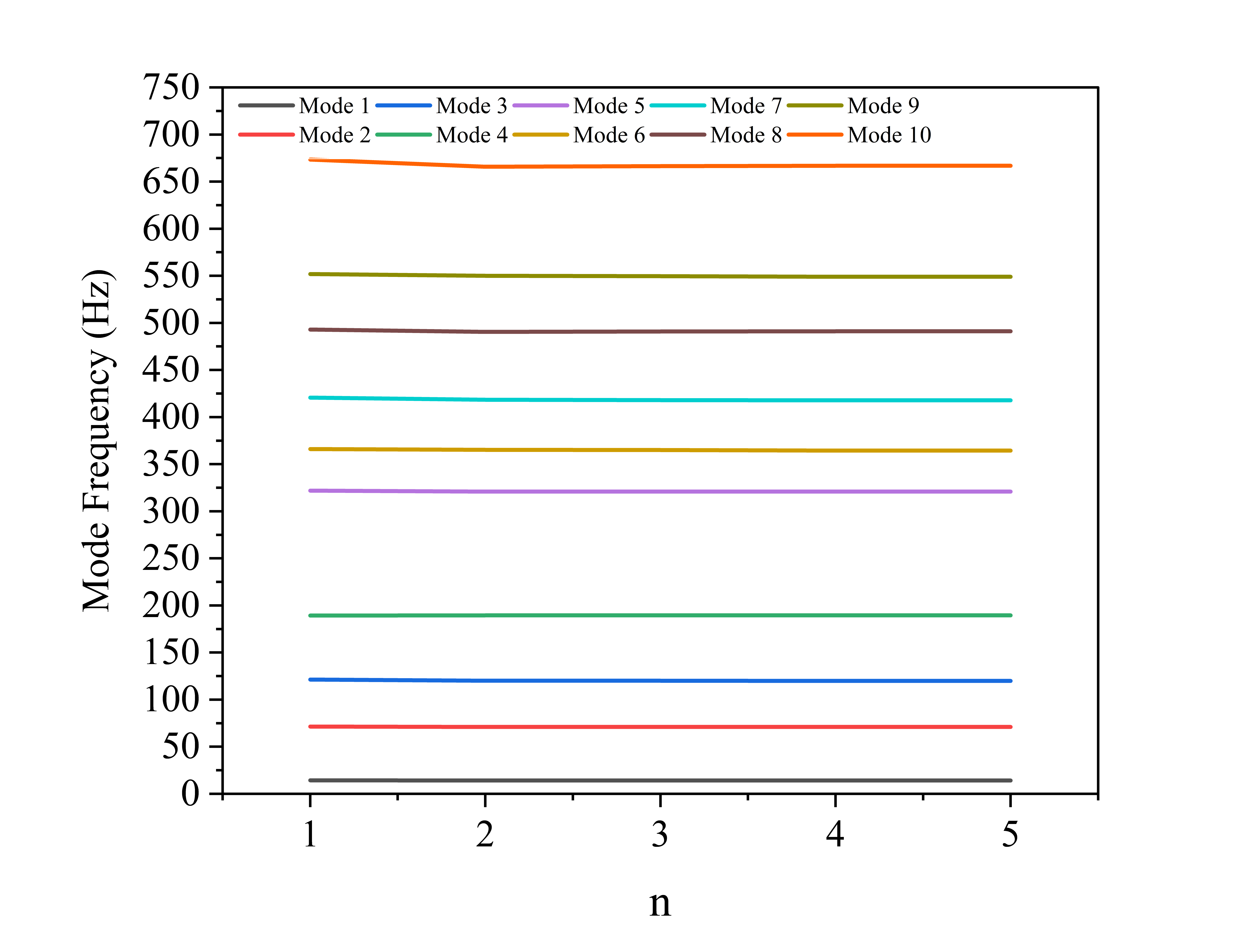}
\end{center}
\caption{Mesh sensitivity analysis.}
\label{fig:fig7}
\end{figure}

The plate is modelled in Nastran using laminate elements for structural representation. A clamped boundary condition is applied at the root of the plate using nodal constraints. A structural mesh of (36×144) is selected after conducting a mesh convergence analysis where the natural frequencies of the first ten modes are determined for different mesh sizes.  Figure \ref{fig:fig7}, where n represents the multiple of 12  chordwise and 48 spanwise elements (12n×48n), respectively, concludes that the analysis is completely mesh-insensitive. 
\subsection{Aerodynamic Model}
Aerodynamic elements are regions of lifting surfaces and are represented using DLM. The rectangular aerodynamic coordinate system defines the flow direction as positive along the x-axis. The aerodynamic mesh is composed of 24×96 elements. DLM is based on the concept of representing a lifting surface as a combination of doublets. By discretizing the lifting surface into a lattice of doublets, the aerodynamic characteristics of the surface can be analysed and calculated \cite{ref21}.      

\subsection{Aeroelastic Equations of motion}
In general, the equations of motion used to describe the behaviour of an aeroelastic system can be expressed in a matrix form: 
\begin{equation}
   \mathbf{A}\ddot{\mathbf{q}} +(\rho V\mathbf{B}+\mathbf{D})\dot{\mathbf{q}} + (\rho V^2\mathbf{C}+\mathbf{E})\mathbf{q} = \mathbf{f}
\end{equation}where \textbf{A} is the structural inertia, \textbf{B} and \textbf{D} are the aerodynamic and structural damping respectively, and \textbf{C} and \textbf{E} are the aerodynamic and structural stiffness respectively. \textbf{q} is the generalised plunge and pitch degrees of freedom. \textbf{f} is the generalized force. V and $\rho$ are the airspeed and air density respectively  \cite{ref22}. In Nastran, a flutter analysis, which determines the dynamic stability of an aeroelastic system, is conducted to find both flutter and divergence speeds. The speeds are found using the British PK method. The primary advantages of the PK method are twofold. Firstly, it directly yields results corresponding to specific velocity values, allowing for precise analysis. Secondly, it offers a reliable estimation of system damping at subcritical speeds, which can be utilized for monitoring flight flutter tests. In the PK method, the aerodynamic matrices are treated as real springs and dampers that depend on frequency. The method involves estimating a frequency and then finding the eigenvalues. From an eigenvalue, a new frequency is obtained. The input data for the PK method allows for looping, enabling the analysis to be performed iteratively. The inner loop contains the velocity set, while the outer loops consider Mach number and density values. This allows for the examination of the effects resulting from changes in one or both parameters within a single run.\\

\subsection{Structure and Aerodynamics Interconnection (Spline Interpolation)}
\begin{figure}[H]
    \centering
    \includegraphics[scale = 0.7]{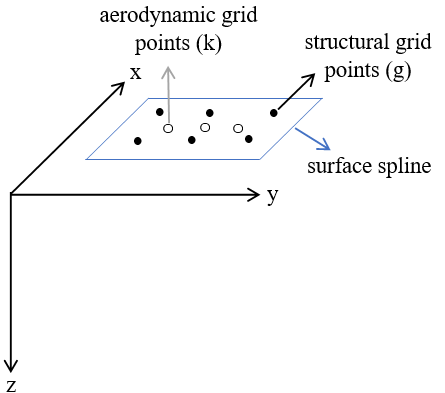}
    \caption{Surface spline and its coordinates (Adapted from \cite{ref23}).}
    \label{fig:fig8}
\end{figure}

For aeroelastic analysis, the aerodynamic (dependent degrees of freedom) and structural meshes (independent degrees of freedom) are connected through splines as shown in Figure \ref{fig:fig8}. Equation \eqref{eqn4} is used to relate the independent and dependent degrees of freedom. Through splining, two transformations are required: an interpolation from the structural to the aerodynamic deflections and a connection between the aerodynamic and the structurally equivalent forces acting on the structural grids. 
\begin{equation}\label{eqn4}
    \{\mathrm{u_K}\} = [\mathrm{G_{kg}}]\{\mathrm{u_g}\}
\end{equation}
where $\mathrm{[G_{kg}}]$ is an interpolation matrix that relates the components of structural grid point deflections $\{\mathrm{u_g}\}$ to the deflections of the aerodynamic grid points $\{\mathrm{u_K}\}$. The interpolation matrix depends on the type of spline used, (surface spline) for this work. The aerodynamic forces, denoted by $\{\mathrm{F_k\}}$ and their corresponding structurally equivalent values $\{\mathrm{F_g\}}$ apply identical virtual work on the structural grid points within their respective deflection modes.
\begin{equation}\label{eqn5}
    \{\mathrm{\delta u_k\}^T}\{\mathrm{F_k\}} = \{\mathrm{\delta u_g\}^T}\{\mathrm{F_g\}}
\end{equation}
where $\mathrm{\delta u_k}$ and $\mathrm{\delta u_g}$ are the aerodynamic and structural virtual deflections respectively, substitute \eqref{eqn4} into \eqref{eqn5} and rearrange to get the required force transformation,
\begin{equation}\label{eqn6}
    \{\mathrm{F_g}\} = \mathrm{[G_{kg}]^T}\{\mathrm{F_k}\}
\end{equation}
Equations \eqref{eqn4} and \eqref{eqn6} are both required to connect aerodynamics and structural grids for any aeroelastic problem \cite{ref23}. 
\subsection{Validation of the baseline plate model}
The baseline plate model is validated by conducting flutter analysis using Nastran. The resulting flutter, divergence, and natural mode frequencies are compared to those obtained from an optimisation study referenced as \cite{ref19}, as well as the experimental and computational work in \cite{ref18} and tabulated in Table \ref{tab:tab3}. 


\begin{table}[h]

  \caption{Baseline plate model validation}
  
  \resizebox{6.7in}{!}{
   \centering
\hskip-4.5cm   
\begin{tabular}{c|c|c|c|c|c|c|c}

    \hline\hline
         Laminate stacking sequence & Fibre Orientation& & Flutter speed(m/s) & Divergence speed(m/s) & \multicolumn{3}{c}{First 3 modes natural frequencies (Hz)} \\\hline \hline
         \multirow{3}*{$[{0^\circ}_2,90^\circ]_s$}&\multirow{3}*{$0^\circ$ spanwise} & Present work & 25 & - & 11 & 39 &69\\

         & & Ref \cite{ref18}(Experimental) & 25 & -& 11 & 42 & 69\\
         & & Ref \cite{ref18}(Computational) & 21 & - & 10.7 & 39 & 67\\
         \hline
         \multirow{2}*{$[-50.7^\circ, 43.2^\circ, 39.2^\circ]_s$}& \multirow{2}*{$0^\circ$ chordwise}& Present work & $\approx 46$ & \multirow{2}{*}{\begin{tabular}[c]{@{}l@{}}(1B) becomes non-oscillatory \\   but does not diverge\end{tabular}}& 5.7 & 35.7 & 75\\
         & &Ref. \cite{ref19} (Computational) & 45.9 & & $\approx 6$ & $\approx 36$& $\approx 75$\\\hline \hline
         
    \end{tabular}
  
    }

    \label{tab:tab3}
\end{table}

It should be noted that the fibre convention utilized in \cite{ref18} was $0^\circ$ spanwise, therefore, to compare their plate with the plate used in this work, the fibre orientation is set to $0^\circ$ spanwise, and the stacking sequence is oriented $90^\circ$.

\section{Parametric Studies and Results Discussion}

Through a series of flutter (SOL 145) and static aeroelastic analysis (SOL 144), this section evaluates the influence of material (mainly ply orientation) and geometry on aeroelastic performance. Two configurations are considered. The first configuration consists of an unstiffened plate, and the plate sweep angle is varied with the ply orientation. The second configuration consists of a stiffened plate where the plate sweep angle, stiffener’s cross-section (I-shaped and T-shaped), and stiffener’s sweep angle are varied with the ply orientation. 

\subsection{Effect of plate sweep angle and ply orientation (Configuration 1)}

The well-known behaviour related to the impact of sweep angle subjected to upwards bending states that; for the upswept configuration, the incidence angle remains unchanged, leading to pure deformation of the plate. For the Aft swept case, the washout effect causes a reduction in the effective streamwise angle of incidence. As a result, the divergence speed increases. Conversely, for the Fw swept case, the washin effect takes place, leading to an increase in the effective streamwise angle of incidence. Consequently, the divergence speed decreases \cite{ref22}. When composites are introduced, the alteration in the geometric tailoring effect due to the sweep angle can be balanced, increased, or decreased.\\


Theta (ply angle) is varied for the range of $-90^\circ<\theta<90^\circ$ with a step of $5^\circ$. As theta is the only design variable, the obtained results directly correspond to the coupling terms of the \textbf{ABD}, specifically the $\textrm{D}_{26}$ (spanwise bend-twist coupling) term. Figure \ref{fig:fig6} illustrates all \textbf{D} components across the entire theta range. Based on the sign of the bending-torsion coupling term, three distinct scenarios can be observed: 1- zero coupling terms, creating a pure bending deflection, 2- negative coupling terms (washout), and 3- positive coupling terms (washin). \\

\begin{figure}[h]
     \centering
     \begin{subfigure}[b]{0.45\textwidth}
         \centering
         \includegraphics[width=\textwidth]{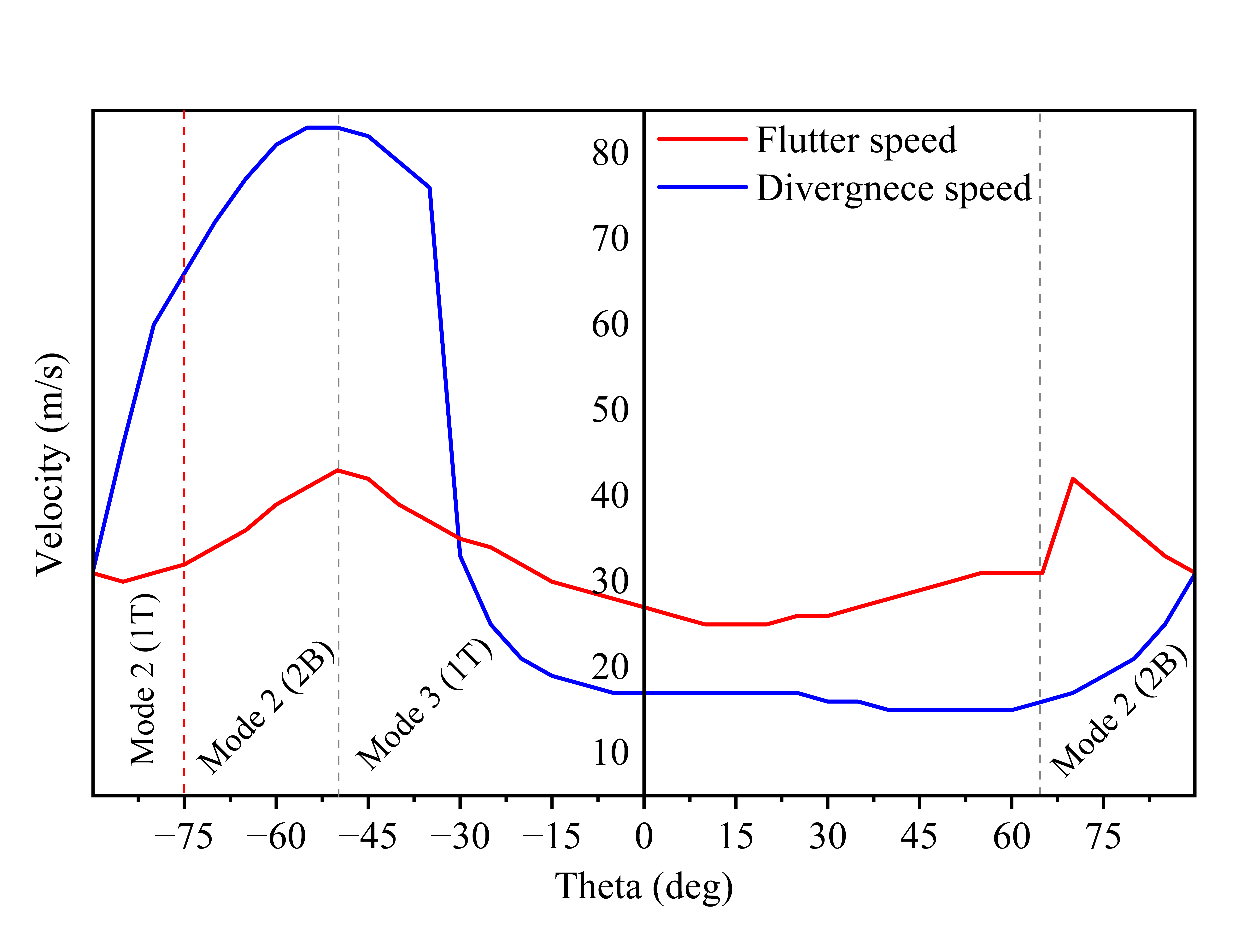}
         \caption{}
         \label{fig:fig9a}
     \end{subfigure}
     \hfill
     \begin{subfigure}[b]{0.45\textwidth}
         \centering
         \includegraphics[width=\textwidth]{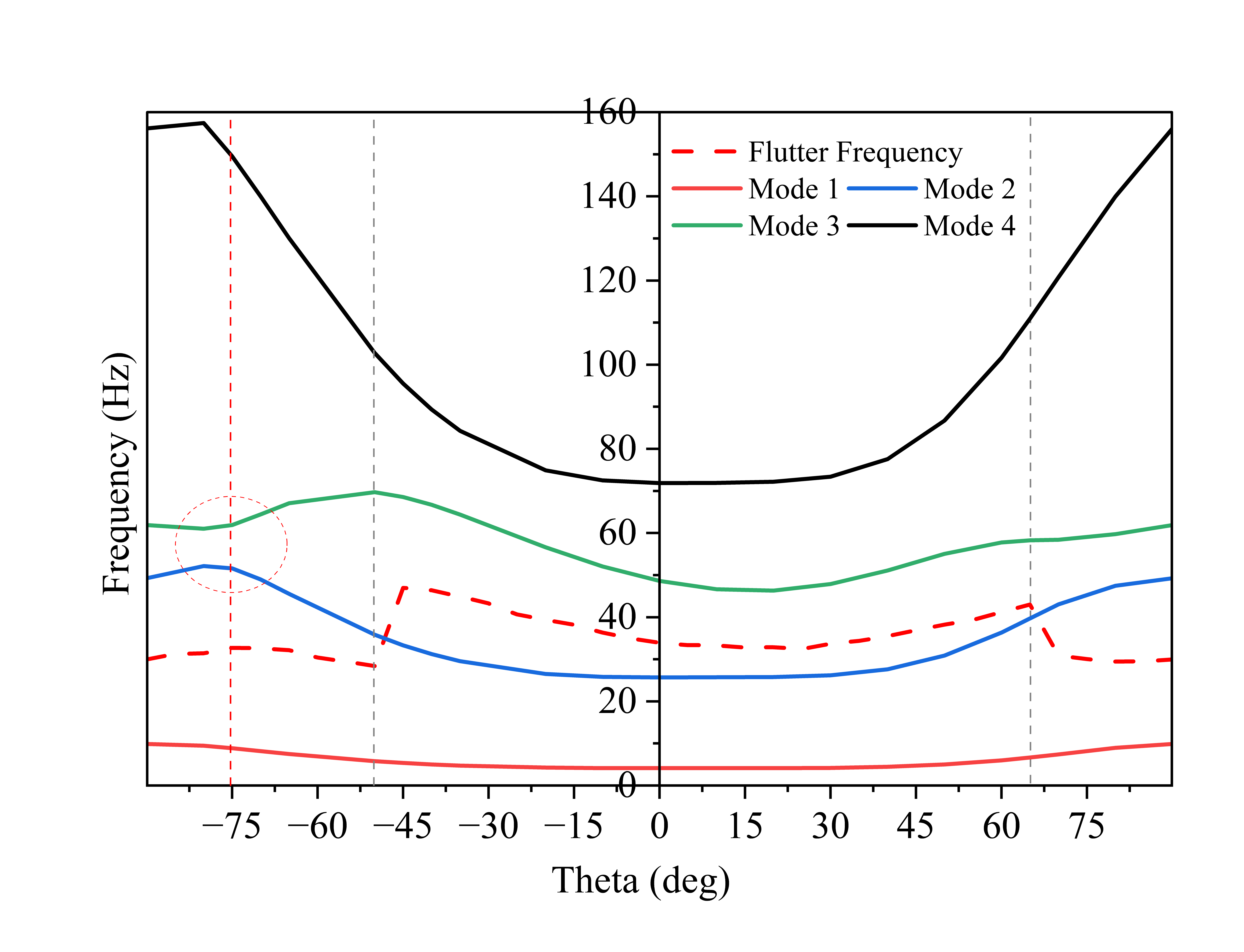}
         \caption{}
         \label{fig:fig9b}
     \end{subfigure}
     \hfill
        \centering
        \caption{(a) Aeroelastic instability velocities vs. theta (b) Flutter and the natural frequencies of the first four modes vs. theta for unswept composite plate.}
        \label{fig:fig9}
\end{figure}

In Figure \ref{fig:fig9a}, the unswept plate exhibits two changes in the flutter mode: at $-50^\circ$ and $65^\circ$, both coinciding with peaks in flutter frequency. Within the range from $-90^\circ$ to $-50^\circ$, the flutter occurs in the second mode, in this region the deformation type changes from first torsion (1T) to second bending (2B) around $-75^\circ$, it can be observed from the frequency Figure \ref{fig:fig9b}  that this transition is accompanied by a convergence between the second and third modes, as these modes approach each other in terms of their frequencies. 

When the fibre orientation is modified, the primary stiffness axis shifts either Aft or Fw. By varying the fibre angle, it is possible that some fibres do not extend continuously from the root to the tip of the plate. This can lead to an unfavourable positioning of the primary stiffness axis. Furthermore, aligning the fibres at a specific angle, such as $-50^\circ$, can be advantageous for both flutter and divergence. Aeroelastic instabilities reach their maximum values at this angle. However, it should be noted that at $-50^\circ$, the fibre orientation introduces a washout effect, which is generally unfavourable for flutter. The peak in divergence speed observed between $-85^\circ$ and $-20^\circ$ is linked to negative coupling terms (washout). Nevertheless, Increasing the divergence speed can be associated with a flutter speed reduction or improvement. Inversely, an increase in flutter may lead to a lower or higher divergence speed; therefore, more design variables are employed in the coming studies to find the best combination of structural and material that fulfils aeroelastic tailoring objectives.

\begin{figure}[h]
\begin{center}
\includegraphics[width=3.7in,height=1.3in]{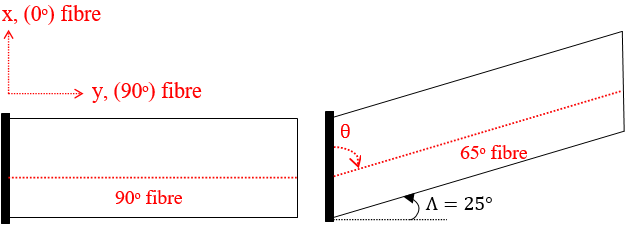}
\end{center}
\caption{Fibre orientation for an unswept and Aft swept plate.}
\label{fig:fig10}
\end{figure}

As stated in the preceding section, the zero degrees fibre is oriented parallel to the velocity vector. In the case of a swept geometry, the convention remains fixed, but it should be noted that fibre orientation is shifted by $25^\circ$ due to the geometrical sweep as illustrated in Figure \ref{fig:fig10}. Furthermore, along the span, if $90^\circ$ is the angle at which the fibres are extended continuously from root to tip, the equivalent would be $65^\circ$ on an Aft swept plate.

\begin{figure}[H]
     \centering
     \begin{subfigure}[b]{0.45\textwidth}
         \centering
         \includegraphics[width=\textwidth]{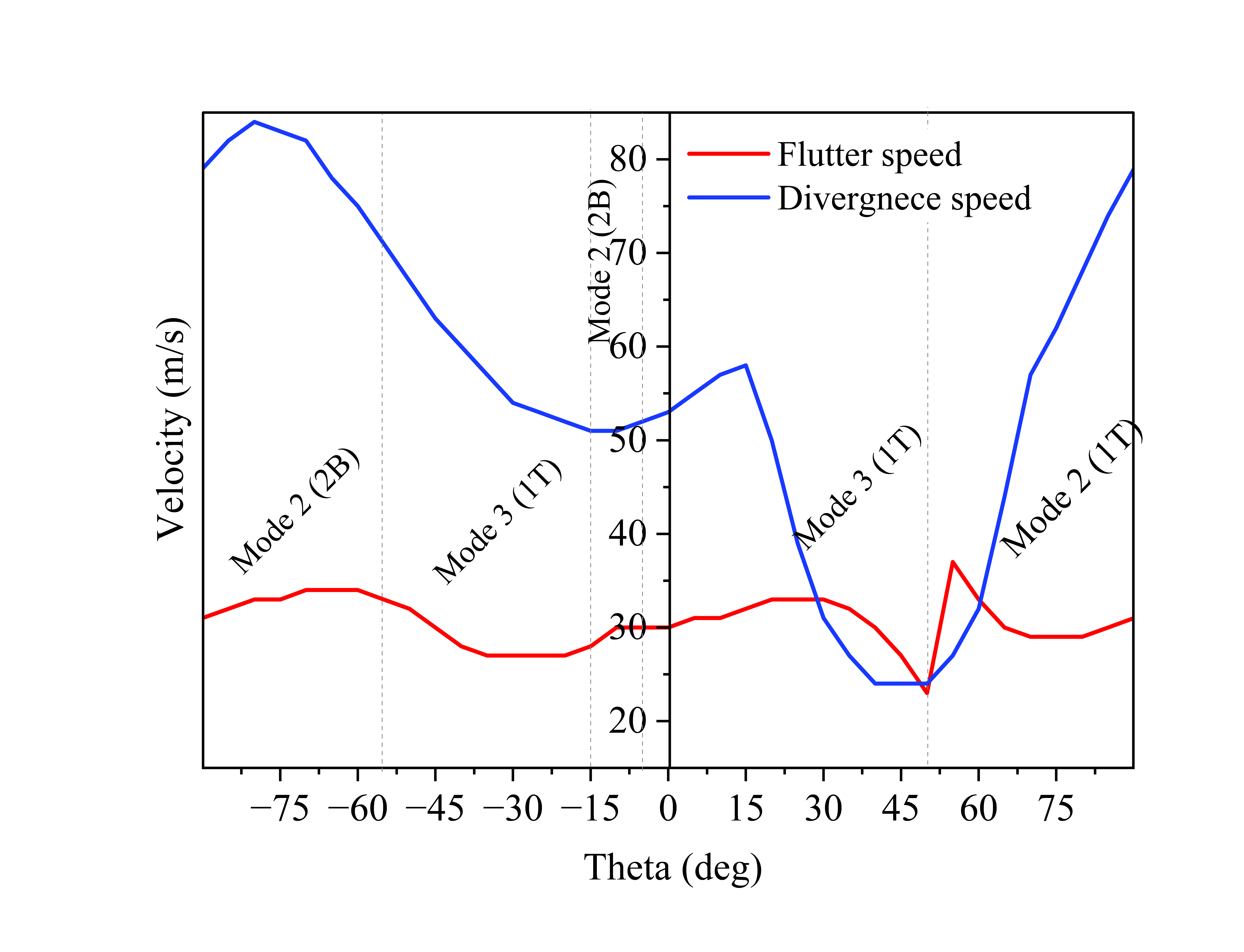}
         \caption{}
         \label{fig:fig11a}
     \end{subfigure}
     \hfill
     \begin{subfigure}[b]{0.45\textwidth}
         \centering
         \includegraphics[width=\textwidth]{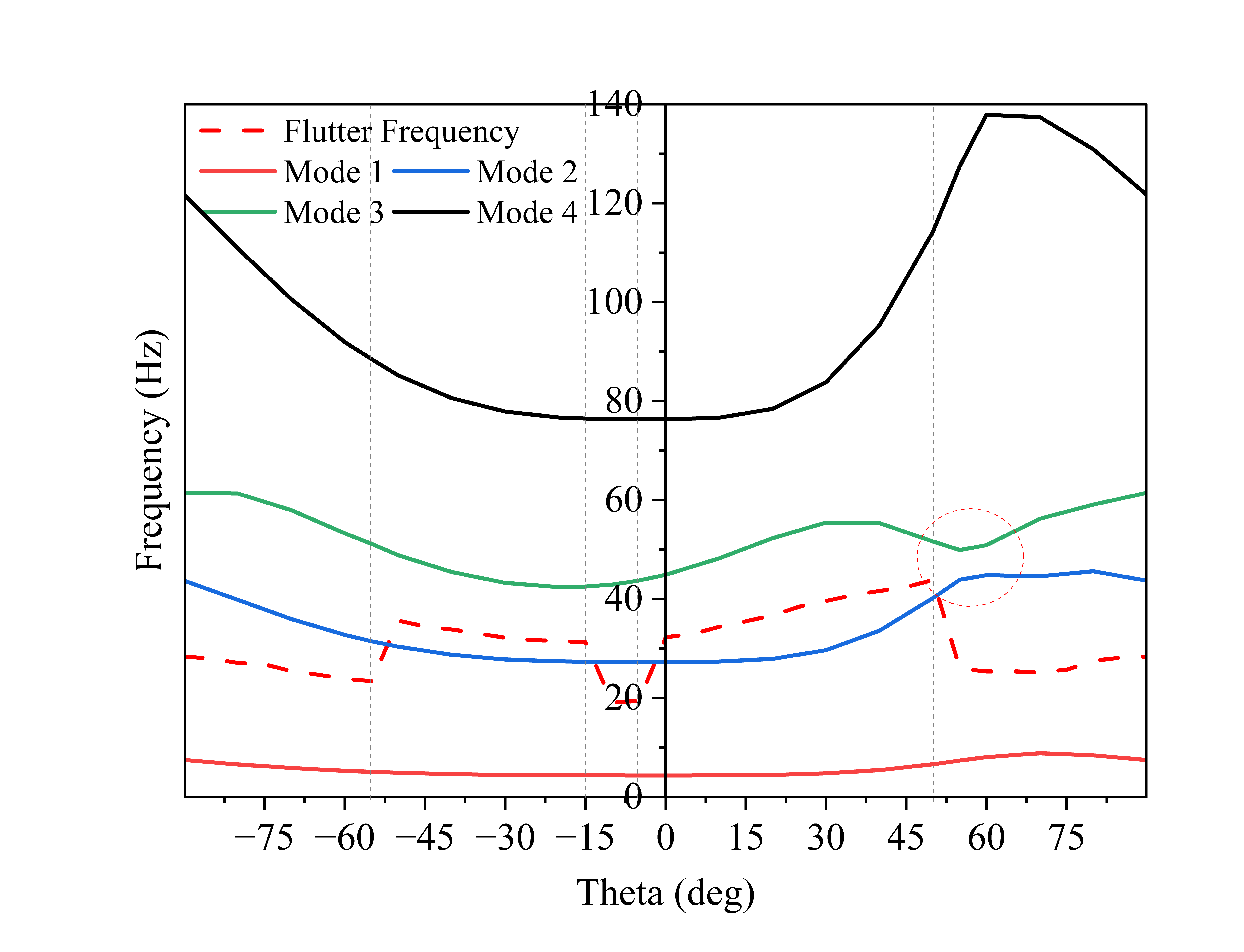}
         \caption{}
         \label{fig:fig11b}
     \end{subfigure}
     \hfill
        \centering
        \caption{(a) Aeroelastic instability velocities vs. theta (b) Flutter and the natural frequencies of the first four modes vs. theta for $25^\circ$ Aft swept composite plate.}
        \label{fig:fig11}
\end{figure}

Figure \ref{fig:fig11} demonstrates the impact of sweeping the plate $25^\circ$ Aft. It is observed that the divergence speed significantly increases compared to the unswept case. However, in the positive region of theta, a substantial decrease in divergence speed is observed. Despite the material’s positive coupling (washin) in this region, the Aft swept geometry (washout) still results in a higher divergence speed compared to the previous case. The figure also shows four changes in the flutter mode, each associated with peaks in flutter frequency. At $50^\circ$, the deformation type of the second mode changes from (1T) to (2B). Alike the previous case, this change coincides with a convergence between the second and third modes as they approach each other in terms of their frequencies.
The maximum flutter speed observed was 37 m/s at $55^\circ$. This is roughly due to the arrangement of the fibres on the swept geometry around this angle. In this region, the fibres are extend continuously from the root to the tip of the plate. Predictably, giving the maximum washin effect on a $25^\circ$ Aft swept geometry. 

\begin{figure}[h]
     \centering
     \begin{subfigure}[b]{0.45\textwidth}
         \centering
         \includegraphics[width=\textwidth]{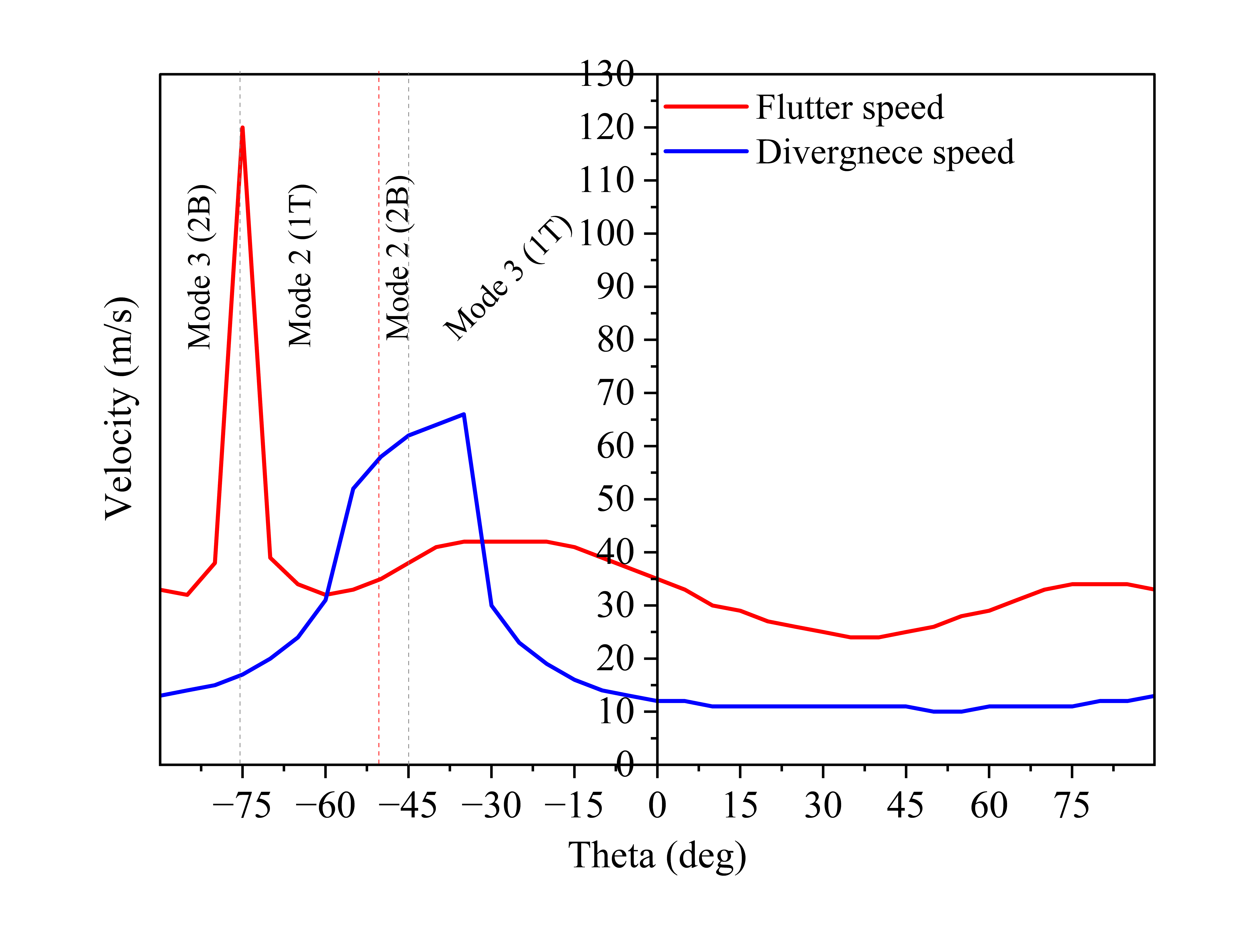}
         \caption{}
         \label{fig:fig12a}
     \end{subfigure}
     \hfill
     \begin{subfigure}[b]{0.45\textwidth}
         \centering
         \includegraphics[width=\textwidth]{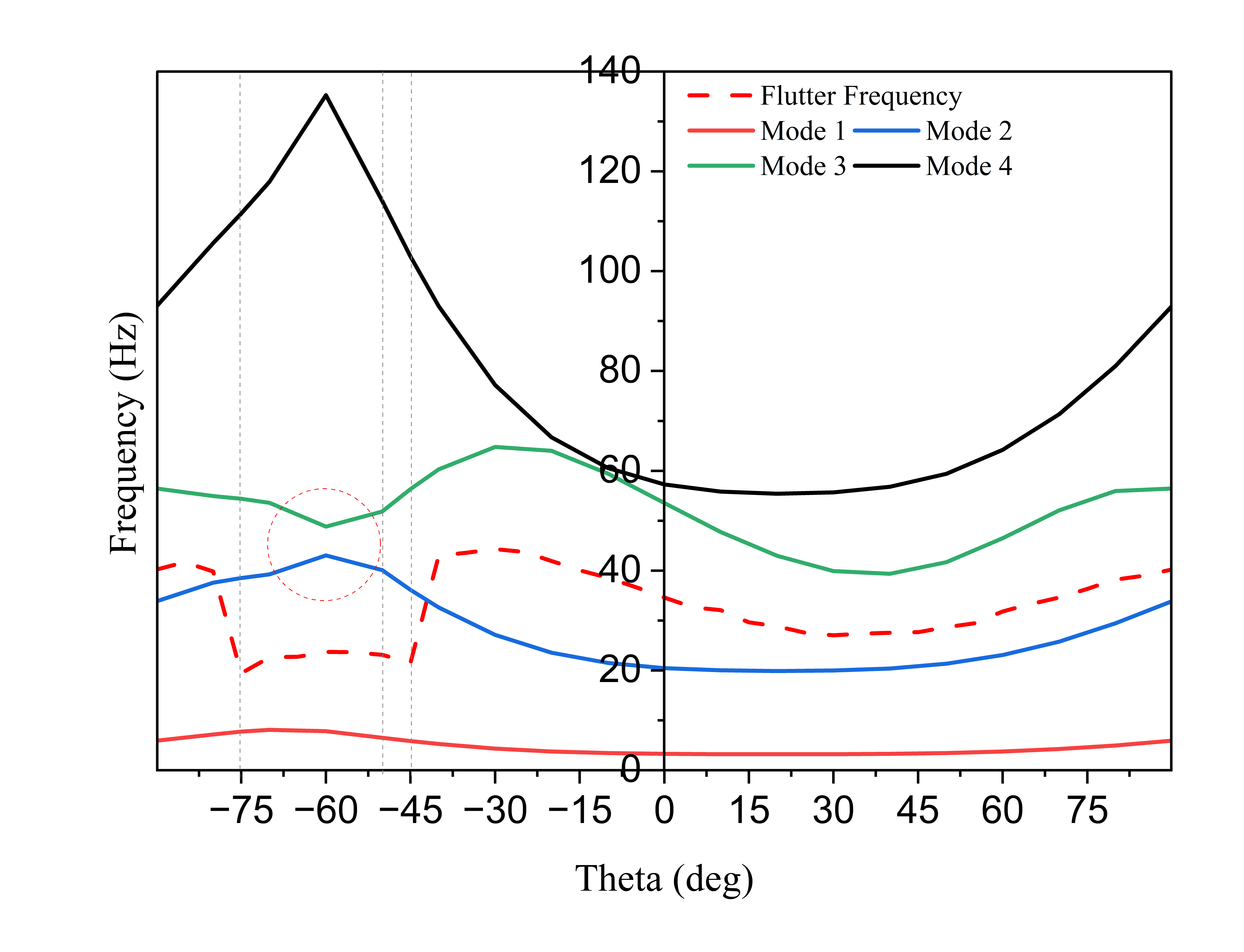}
         \caption{}
         \label{fig:fig12b}
     \end{subfigure}
     \hfill
        \centering
        \caption{(a) Aeroelastic instability velocities vs. theta (b) Flutter and the natural frequencies of the first four modes vs. theta for $25^\circ$ Fw swept composite plate.}
        \label{fig:fig12}
\end{figure}

Figure \ref{fig:fig12} shows that sweeping the plate $25^\circ$ Fw results in an improvement in flutter speed when compared with the Aft sweep. However, as a consequence of this flutter increase, the divergence occurs at very low speeds, except in the region where the material exhibits a washout effect (in a sense; balancing the effect of geometry). Between $-75^\circ$ and $-45^\circ$ the flutter occurs in the second mode, and within this region, the deformation type changes from (1T) to (2B), where the modes approach each other in terms of their frequencies. 

In summary, a maximum flutter speed of 120 m/s, which occurs in the third mode (2B), is achieved when sweeping the plate $25^\circ$ Fw and orienting the fibres by $-75^\circ$. On the other hand, the optimal divergence speed is 84 m/s, which occurs in the first mode (1B), when sweeping the plate $25^\circ$ Aft and orienting the fibres by $-80^\circ$. These findings highlight the impact of both sweep direction and fibre orientation on the aeroelastic behaviour of the plate, with different combinations yielding distinct flutter and divergence characteristics.
\subsection{Effect of stiffening the plate (Configuration 2 with unswept stringers)}
\begin{figure}[h]
     \centering
     \begin{subfigure}[b]{0.45\textwidth}
         \centering
         \includegraphics[width=\textwidth]{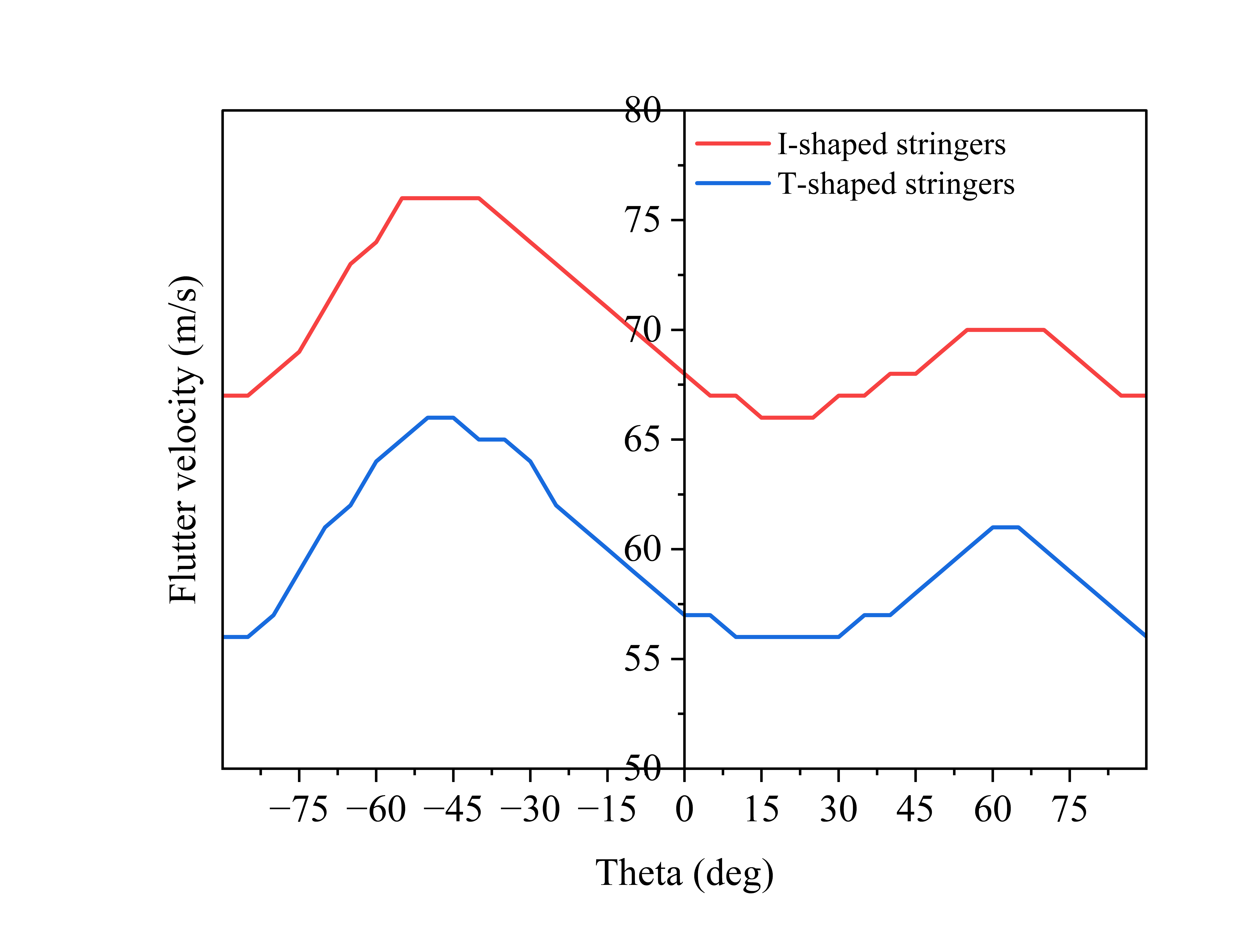}
         \caption{}
         \label{fig:fig13a}
     \end{subfigure}
     \hfill
     \begin{subfigure}[b]{0.45\textwidth}
         \centering
         \includegraphics[width=\textwidth]{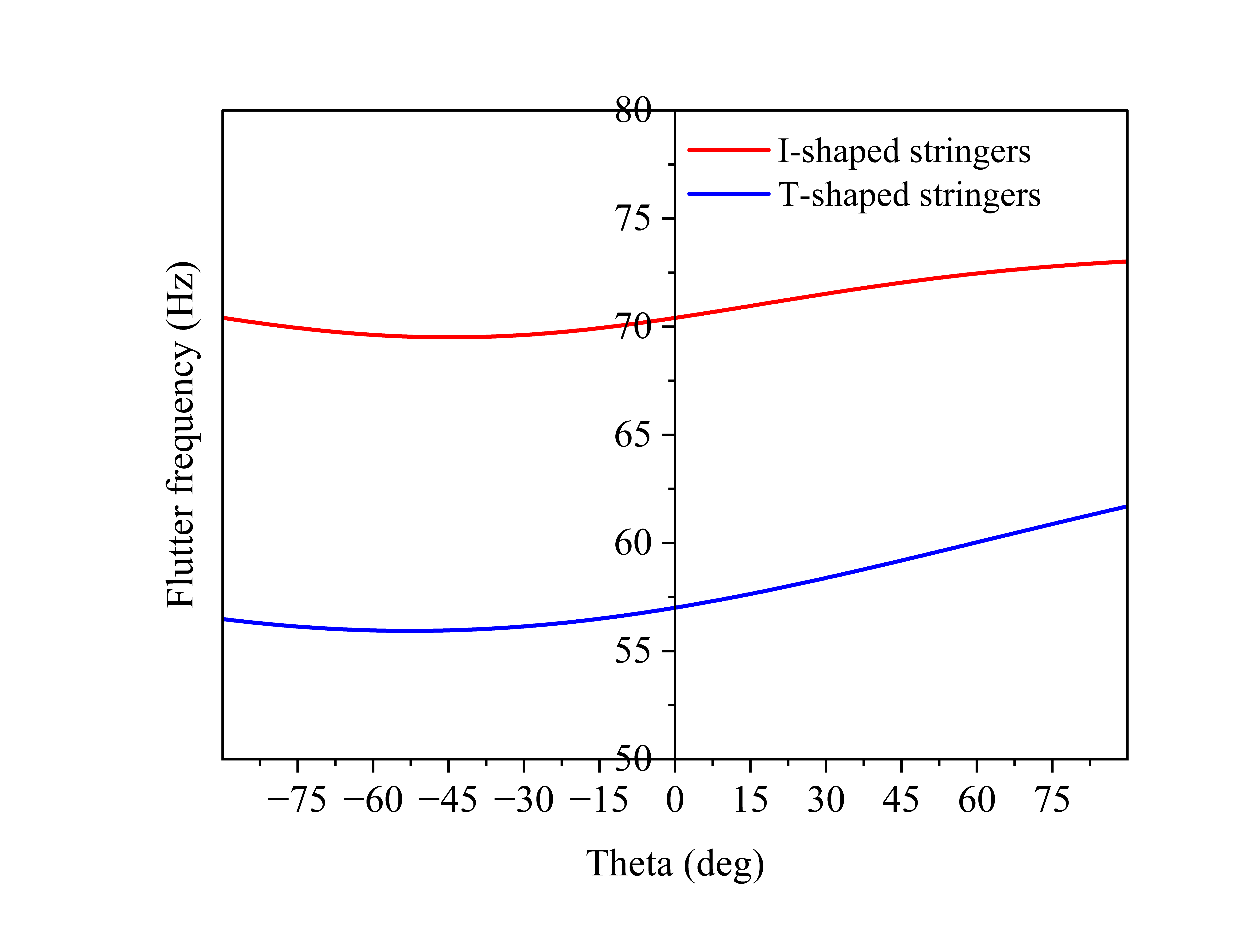}
         \caption{}
         \label{fig:fig13b}
     \end{subfigure}
     \hfill
        \centering
        \caption{Flutter speed vs. theta (b) Flutter frequency vs. theta for the unswept stiffened plate with I-shaped and T-shaped stringers.}
        \label{fig:fig13}
\end{figure}

\begin{figure}[H]
    \centering
    \includegraphics[width=2.25in,height=1.7in]{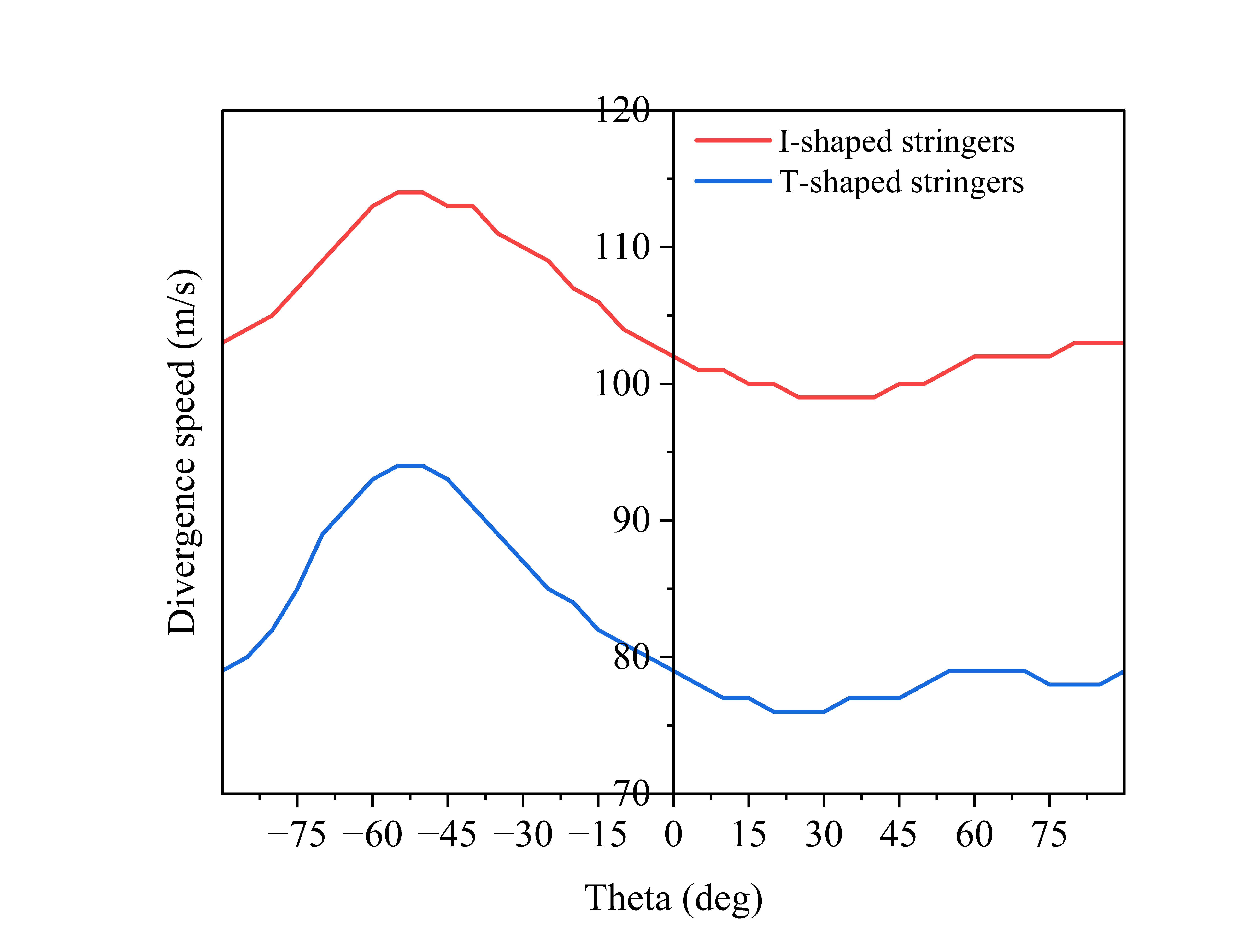}
    \caption{Divergence speed vs. theta for the unswept stiffened plate with I-shaped and T-shaped stringers.}
    \label{fig:fig14}
\end{figure}

In this section, the focus is on investigating the influence of incorporating stiffeners into an unswept composite plate, as well as assessing the effects of altering the cross-sectional shape of the stringers. Both I and T-shaped stringers are considered. As shown in Figure \ref{fig:fig13a} and Figure \ref{fig:fig14}, the influence of theta on aeroelastic instabilities is relatively minor. In the negative region of theta, where the coupling term ($\mathrm{D}_{26}$ ) is negative, and the material is giving a washout effect, a slight increase in aeroelastic instabilities is observed compared to the positive range of theta.  This trend is similar to that of the unswept plate (unstiffened) where an increase in flutter and divergence speeds is achieved in the negative range of theta. However, the increase achieved in the negative range of theta for the unswept stiffened plate is reduced. This suggests that the variation in theta has a limited impact on the flutter and divergence characteristics of the stiffened plate mainly because the stiffeners are the main contributors to the rigidity of the structure. Moreover, the behaviour of the first and second modes as well as the flutter frequency (Figure \ref{fig:fig13b} ) is almost independent of theta. 
In contrast to subsection (1) where changes in the flutter mode (between the 2nd and 3rd modes) are observed for different theta values, no changes are observed for the entire range of theta when stiffeners are added. Regardless of the stiffener’s cross-section, the flutter always occurs in the second mode (1T). This further supports the notion that the presence of stringers and the resulting increase in stiffness have a stabilizing effect on the plate's aeroelastic behaviour. 

Both I and T-shaped stringers exhibit a comparable pattern in terms of aeroelastic behaviour. This similarity can be attributed to their relatively similar geometry and mass distribution near the attachment area with the plate. Figure \ref{fig:fig13} and Figure \ref{fig:fig14} illustrate this trend, highlighting the influence of the stringer cross-section shape on aeroelastic performance. Moreover, the utilization of T-shaped stringers results in the vertical positions of the centre of gravity and the effective neutral axis being closer to the plate, compared to I-shaped stringers. On the other hand, the additional flange in the I-shaped stringers shifts the centre of gravity and the neutral axis downwards, providing higher bending rigidity. This causes variations in mode shapes and natural frequencies.

\subsection{Effect of varying the stringer's sweep angle (Configuration 2).}

In this subsection, the influence of sweeping the stringers is examined. It is observed that sweeping the stringers alone has a different effect compared to sweeping the entire plate. Unlike sweeping the entire plate, where Aft sweep results in a washout effect and Fw sweep results in a washin effect, sweeping the stringers alone reverses this well-known behaviour. This observation highlights the importance of considering the specific configuration on a more realistic wing box model. Furthermore, this finding suggests that sweeping the stringers alone could potentially address the aeroelastic instabilities associated with Fw swept wings and provide a solution for such design considerations.

\begin{figure}[H]
    \centering
    \includegraphics[width=3.1in,height=2.25in]{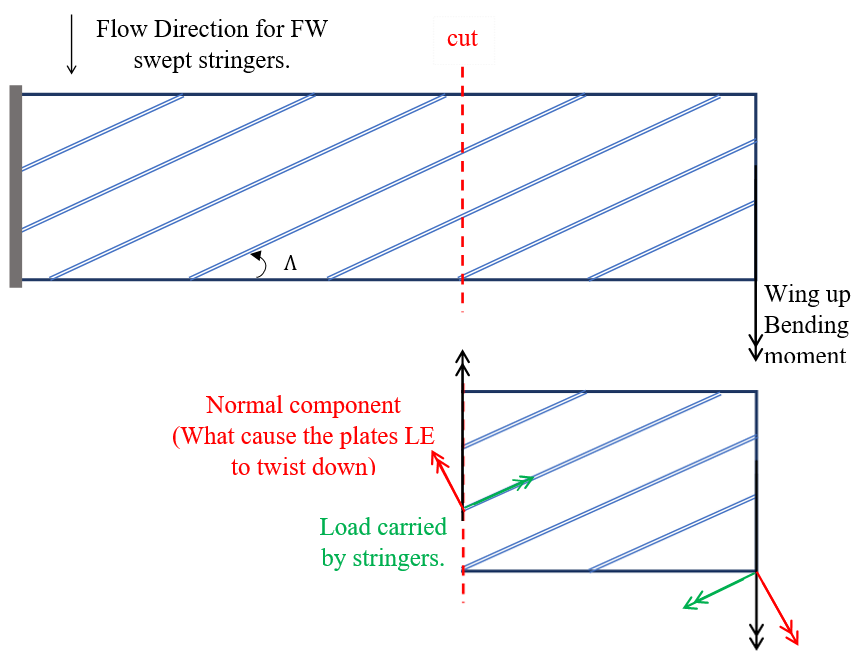}
    \caption{Loads redistribution along the swept stringers.}
    \label{fig:fig15}
\end{figure}

The effect of sweeping the stringers with respect to the applied load axis generates shear loads within the skin. For Fw swept stringers subjected to upward bending (from lift) results in a washout deformation. On the contrary, a plate with Aft swept stringers under upward bending (from lift) experiences a washin deformation.  Figure \ref{fig:fig15} shows how the applied load is resolved into two components, one acting along the stringer’s direction and one normal to the stringer's direction. The normal component is what causes the plate to twist (nose-down).  The coupling effect may differ based on the amount of the stringer’s orientation angle \cite{ref24, ref25}.  
\subsubsection{Static aeroelastic analysis}
A static aeroelastic analysis is performed to assess the influence of stringer's sweep angle on the aeroelastic behavior of the stiffened plate (configuration 2). The symmetric laminate  ([$90^\circ,45^\circ,-45^\circ]_s$) is chosen for the analysis. A range of angle of attack between $0^\circ$ and $5^\circ$ with a step of $1^\circ$ is considered at sea level and an airspeed of 20 m/s.\\

\begin{figure}[h]
     \centering
   \begin{subfigure}[b]{0.45\textwidth}
        \centering
     \includegraphics[width=\textwidth]{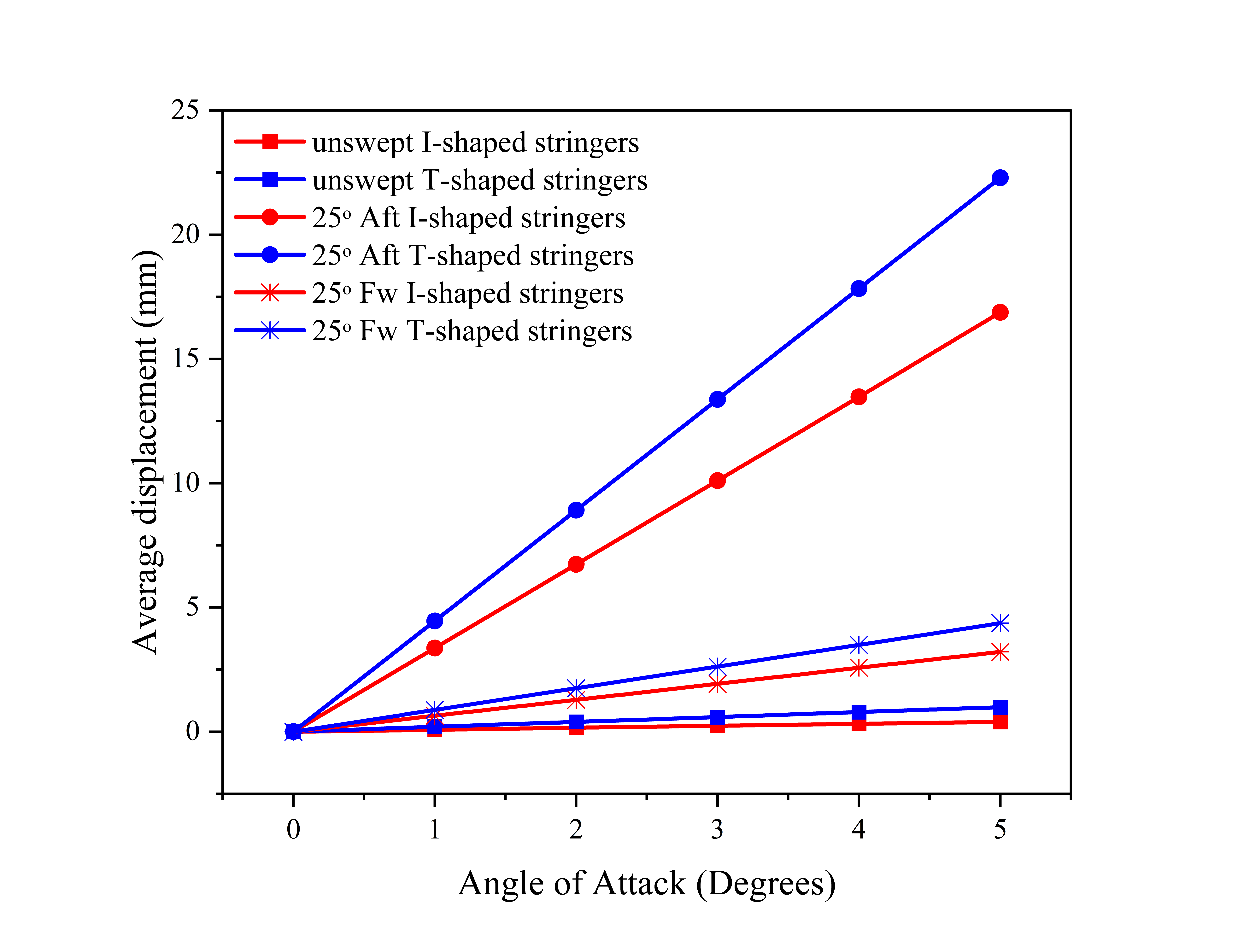}
         \caption{}
         \label{fig:fig20}
    \end{subfigure}
     \hfill
    \begin{subfigure}[b]{0.45\textwidth}
        \centering
         \includegraphics[width=\textwidth]{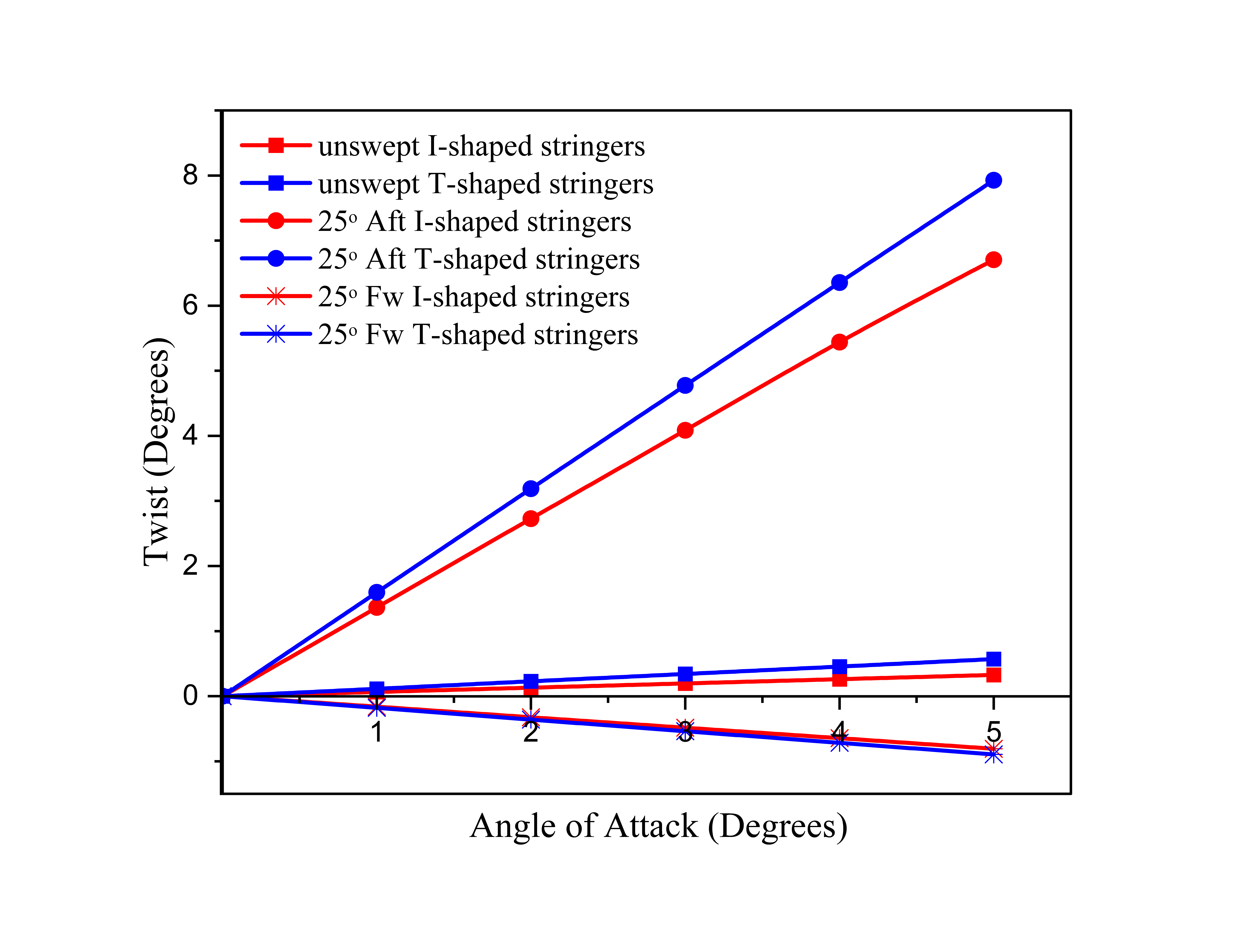}
         \caption{}
         \label{fig:fig21}
     \end{subfigure}
     \hfill
        \centering
        \caption{(a) Average tip displacement vs. angle of attack (b) Tip twist vs. angle of attack for configuration 2.}
        \label{fig:fig21all}
\end{figure}

Figure \ref{fig:fig21all} shows the variation of average tip displacement in the vertical direction (z-axis) and tip twist across the range of angle of attack. 
It can be noted that the average tip displacement and twist will increase with increasing the angle of attack due to the increased aerodynamic forces acting on the plate. However, the tip displacement can be reduced by utilizing Fw swept stringers. This is mainly attributed to the nose-down tip twist that is achieved with Fw swept stringers as illustrated in Figure \ref{fig:fig21}, which reduces the effective streamwise angle of incidence of the plate providing a washout effect. On the contrary, Aft swept stringers result in a nose-up twist, which will result in increasing the aerodynamic loads, hence, increasing the average tip displacement. Moreover, since the selected laminate is giving a washin effect (Figure \ref{fig:fig6}), this also can contribute to the increase in tip average displacement caused by the Aft swept stringers \\  

As stated earlier, the additional flange in the I-shaped stringers shifts the centre of gravity and the effective neutral axis downwards, providing higher bending rigidity, thus, the I-shaped stringers exhibit a less average tip displacement compared to T-shaped stringers. Nevertheless, it can be noted that the effect of the stringer's cross-section is increased when sweeping the stringers Aft.\\

\subsubsection{Flutter Analysis}
\begin{figure}[h]
     \centering
   \begin{subfigure}[b]{0.45\textwidth}
        \centering
     \includegraphics[width=\textwidth]{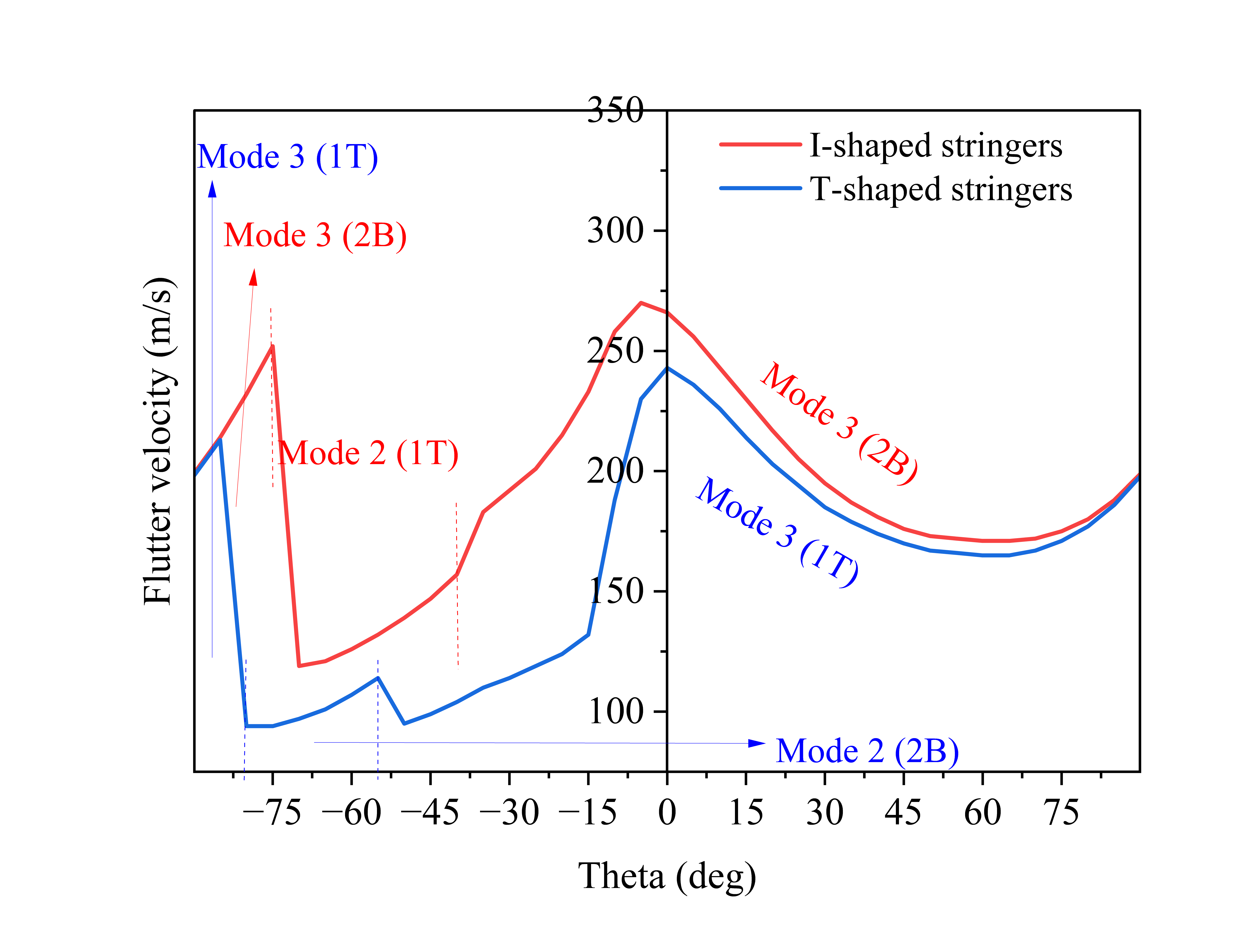}
         \caption{}
         \label{fig:fig16a}
    \end{subfigure}
     \hfill
    \begin{subfigure}[b]{0.45\textwidth}
        \centering
         \includegraphics[width=\textwidth]{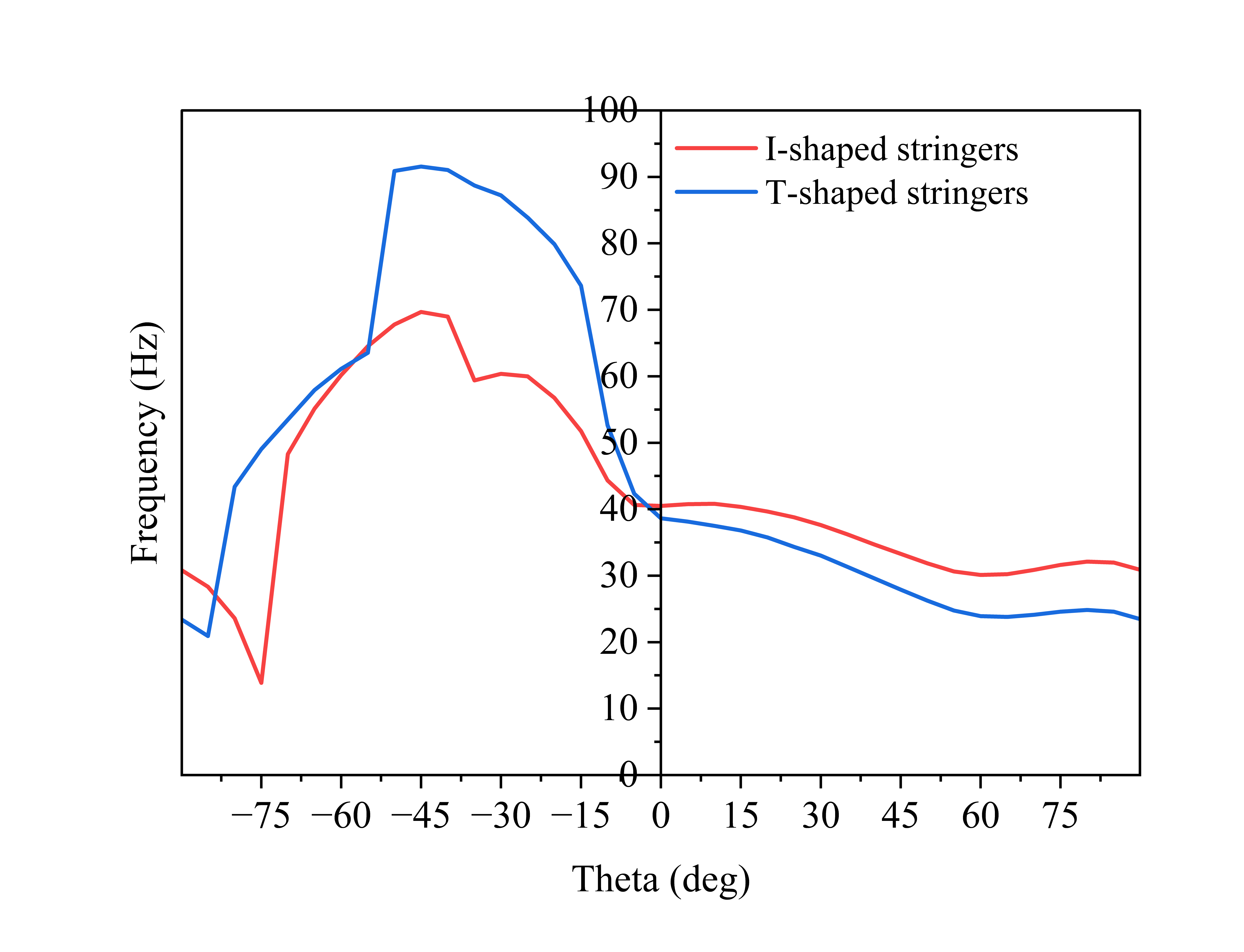}
         \caption{}
         \label{fig:fig16b}
     \end{subfigure}
     \hfill
        \centering
        \caption{(a) Flutter speed vs. theta (b) Flutter frequency vs. theta for the unswept stiffened plate with $25^\circ$ Aft swept I-shaped and T-shaped stringers.}
        \label{fig:fig16}
\end{figure}

\subsubsubsection{Aft swept stringers}
Figure \ref{fig:fig16} illustrates the impact of varying the ply orientation, sweeping the stringers $25^\circ$ Aft, and changing the stringer’s cross-section. Several conclusions can be drawn from Figure \ref{fig:fig16a}. In the positive region of theta, where both the material and geometry contribute to a washin effect, the flutter speed is generally higher. This can be attributed to the combined effect of the favourable geometry and material properties. In the negative range of theta, where the geometry provides a washin effect while the material gives a washout effect, it can be observed that the flutter mode switches and this is associated with frequency peaks as shown in Figure \ref{fig:fig16b}. 
\begin{figure}[H]
    \centering
   \includegraphics[width=2.25in,height=1.7in]{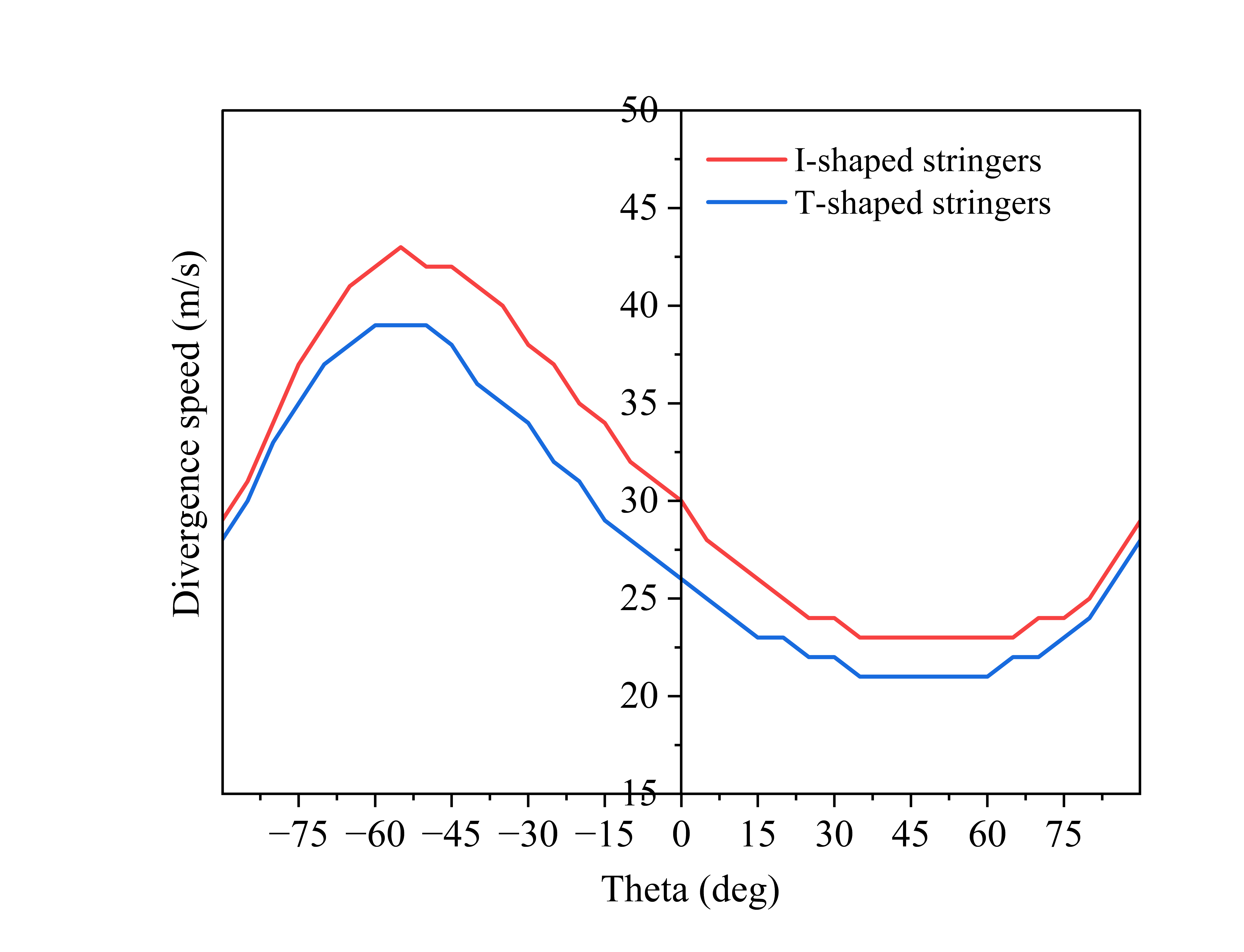}
    \caption{Divergence speed vs. theta for the unswept stiffened plate with $25^\circ$ Aft swept I-shaped and T-shaped stringers.}
    \label{fig:fig17}
\end{figure}
When I-shaped stringers are employed, Figure \ref{fig:fig16a} indicates two changes in the flutter mode between the third and second modes ((2B) and (1T) respectively). These mode changes can be attributed to different factors. The first change in flutter mode may be influenced by the transition in the material effect from washin to washout, as depicted in Figure \ref{fig:fig6}. The second change in flutter mode could be associated with the material effect as well. In this range, an intersection of the $\mathrm{D}_{26}$ and $\mathrm{D}_{16}$ terms occurs (Figure \ref{fig:fig6}).  
Similarly, when T-shaped stringers are employed, two changes in flutter mode can be observed around the same region. These mode changes and the shift in the dominant mode of flutter occur due to similar underlying assumptions, involving the interplay between material effects and geometric factors.

Figure \ref{fig:fig17} demonstrates the influence of the material effect on the divergence speed of the stiffened plate with Aft swept stringers. Specifically, in the negative range of theta, where the material exhibits a washout effect, an increase in divergence speed can be observed regardless of the stringer’s cross-section.

\begin{figure}[h]
     \centering
   \begin{subfigure}[b]{0.45\textwidth}
        \centering
     \includegraphics[width=\textwidth]{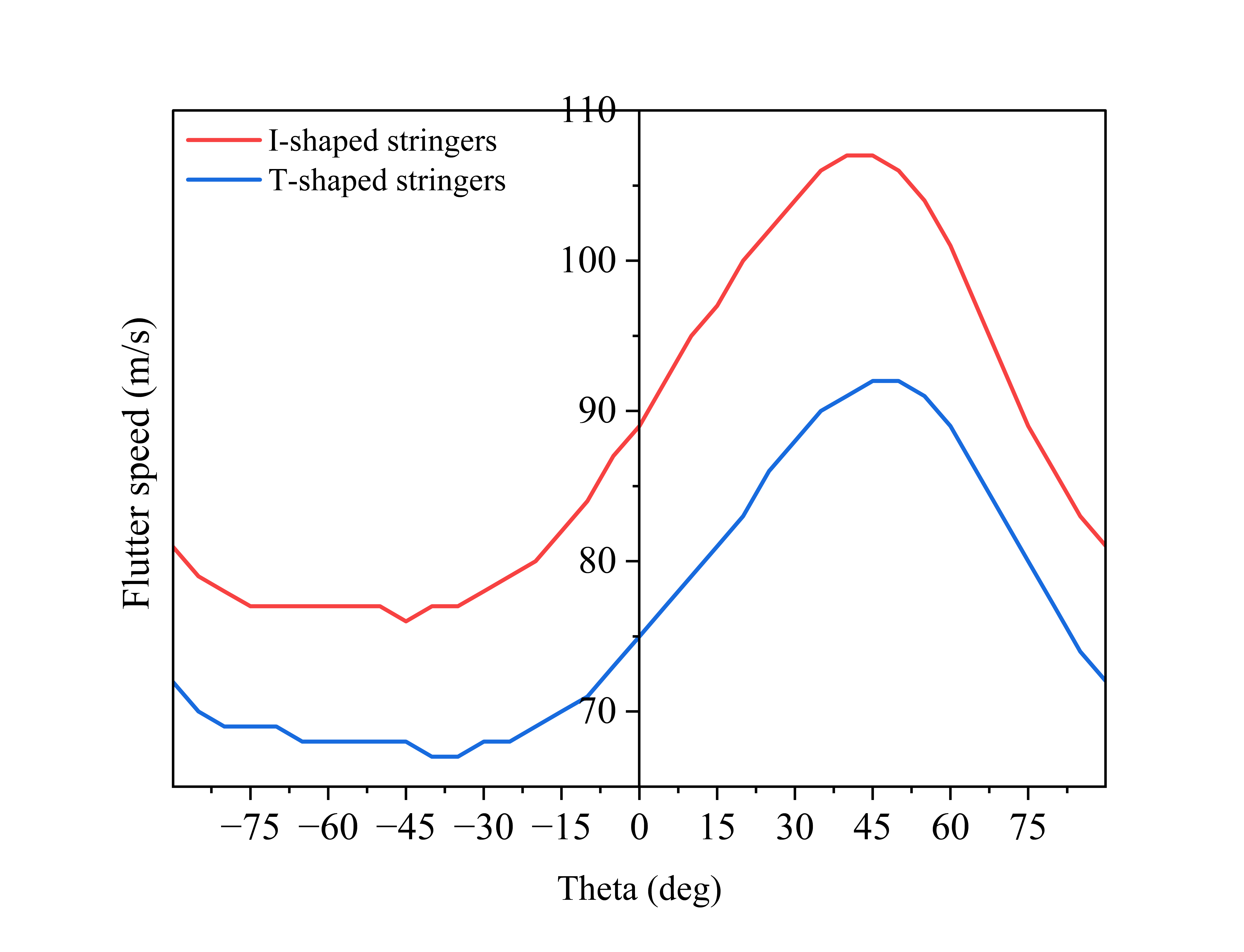}
         \caption{}
         \label{fig:fig18a}
    \end{subfigure}
     \hfill
    \begin{subfigure}[b]{0.45\textwidth}
        \centering
         \includegraphics[width=\textwidth]{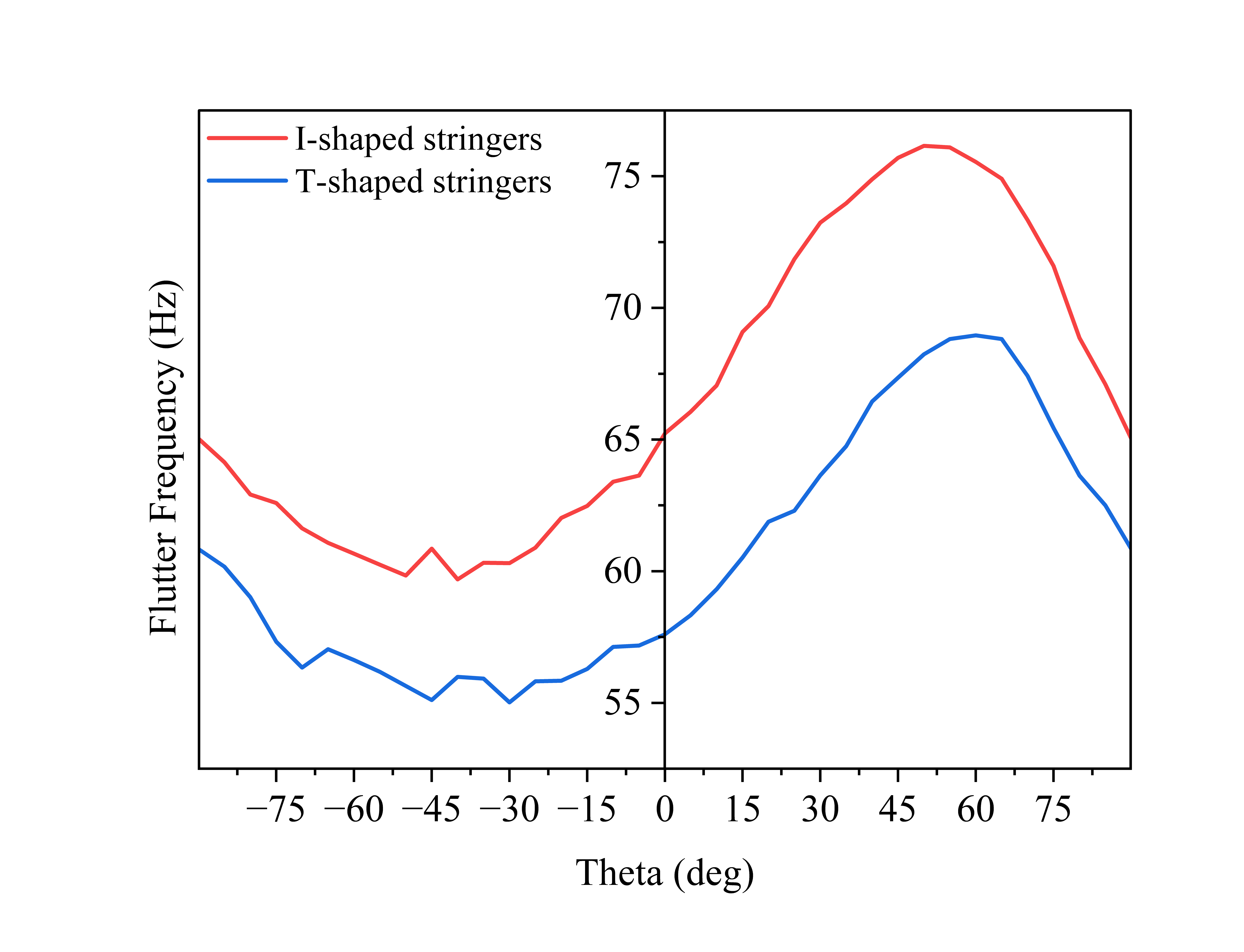}
         \caption{}
         \label{fig:fig18b}
     \end{subfigure}
     \hfill
        \centering
        \caption{(a) Flutter speed vs. theta (b) Flutter frequency vs. theta for the unswept stiffened plate with $25^\circ$ Fw swept I-shaped and T-shaped stringers.}
        \label{fig:fig18}
\end{figure}

 \subsubsubsection{Fw swept stringers.}
 Figure \ref{fig:fig18} illustrates the influence of sweeping the stringers Fw, changing the stringer’s cross-section, and varying the ply orientation on the aeroelastic behaviour of the stiffened plate. It can be observed that the flutter speed for both cross-sections is higher in the positive region of theta, where the material is giving a washin effect, thus balancing the washout effect caused by the geometry. Nevertheless, as concluded earlier, I-shaped stringers result in higher aeroelastic instabilities speeds compared to T-shaped stringers. 
 
\begin{figure}[H]
    \centering
   \includegraphics[width=2.4in,height=1.7in]{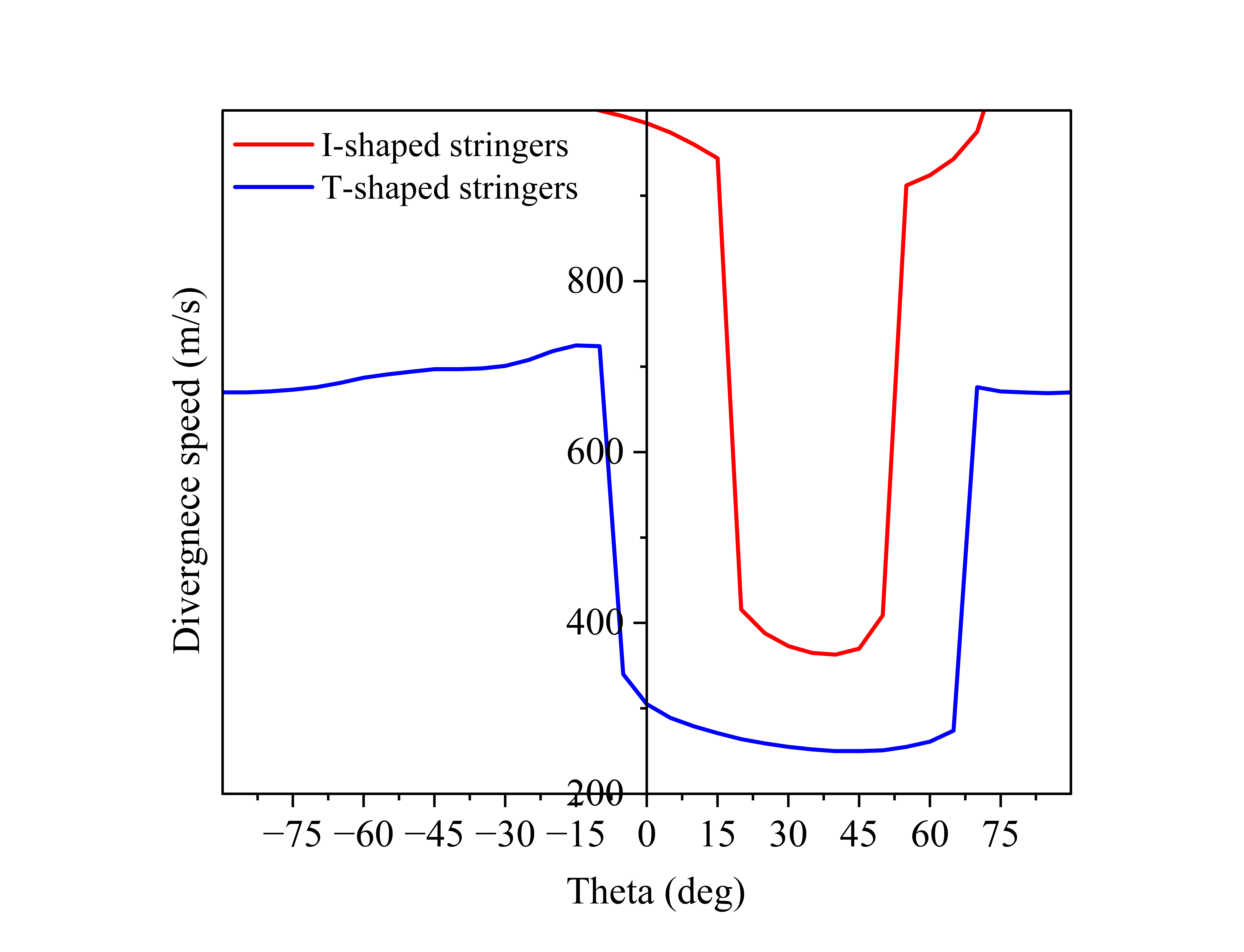}
    \caption{Divergence speed vs. theta for the unswept stiffened plate with $25^\circ$ FW swept I-shaped and T-shaped stringers.}
    \label{fig:fig19}
\end{figure}

 Figure \ref{fig:fig19} illustrates the plate's divergence speed, which is extremely high for both cross-sections. The I-shaped stringers result in a divergence speed above 360 m/s, while the T-shaped stringers exhibited a divergence speed greater than 250 m/s. The significant increase in the divergence occurs when both the material and geometry contribute to the washout effect. From Figure \ref{fig:fig19}, it can be shown that the divergence speed for the plate stiffened with I-shaped stringers cannot be detected in the negative range of theta, as it exceeds 1100 m/s.  These high divergence speeds make the flutter velocity a critical consideration for this particular case. The washout effect caused by the geometry contributes to the high divergence velocities observed in this study. 


\subsection{Conclusions}
This paper investigated the aeroelastic performance of a stiffened composite plate. The study focused on the effects of stiffening the plate with different cross-section stringers and the impact of sweeping the stringers and the plate. The key findings and conclusions from the analysis are as follows:
\begin{Romanlist}
  \item By stiffening the plate, the flutter frequency tends to increase. This indicates that the structure becomes more resistant to aeroelastic instabilities as its natural frequencies increase. 
  \item Sweeping the stringers alone on an unswept plate reversed the well-known direction of the primary stiffness axis. Sweeping the stringers Aft resulted in a washin effect, increasing the flutter speed, while sweeping the stringers Fw caused a washout effect, increasing the divergence speed significantly. 
  \item Varying the ply orientation had a less significant impact on controlling flutter compared to stiffening the plate with different cross-section stringers. The material effect mainly influenced the divergence behaviour.
  \item Significant aeroelastic tailoring can be acheived using Fw swept stringers. The aeroelastic tailoring capability increases with angle of attack and dynamic pressure. It was also noticed that I-shaped stringers are more effective than T-shaped stringers due to their higher bending rigidity creating more favourable bending-twist coupling.

\end{Romanlist}
Overall, the results demonstrated that the choice of stringer cross-section and the geometric configuration of the stiffened plate had a dominant influence on aeroelastic instabilities. The study emphasized the importance of considering both material and geometric factors in designing and analysing stiffened composite plates to mitigate aeroelastic instabilities and ensure structural integrity. \\

\textbf{Competing interests:} The authors declare none.
\bibliographystyle{ieeetr}
\bibliography{ref}
\end{document}